\documentclass[journal]{IEEEtran}

\usepackage{epsfig}
\usepackage[noadjust]{cite}
\usepackage{tikz}
\usepackage{multirow}
\usepackage{caption}
\usepackage{subcaption}
\usepackage[inline]{enumitem}
\usepackage[bookmarks=false]{hyperref}
\usepackage{amsmath}
\usepackage{amssymb}
\usepackage{multicol}
\usepackage{array}
\usepackage{textgreek}
\usepackage{footnote}
\makesavenoteenv{tabular}
\makesavenoteenv{table}
\usepackage{algorithmic}
\usepackage{algorithm}
\usepackage[absolute,showboxes]{textpos}
\floatname{algorithm}{Algorithm}

\usetikzlibrary{shapes.geometric}

\ifCLASSINFOpdf
\else
\fi

\definecolor{gray}{rgb}{0.75, 0.75, 0.75}

\TPMargin{5pt}

\newcommand{\copyrightstatement}{
	\begin{textblock}{0.84}(0.08,0.93)    
		\noindent
		\scriptsize
		\copyright 2020 IEEE. Personal use of this material is permitted. However, permission to reprint/republish this material for advertising or promotional purposes or for creating new collective works for resale or redistribution to servers or lists, or to reuse any copyrighted component of this work in other works, must be obtained from the IEEE. Contact: Manager, Copyrights and Permissions / IEEE Service Center / 445 Hoes Lane / P.O. Box 1331 / Piscataway, NJ 08855-1331, USA. Telephone: + Intl. 908-562-3966.
	\end{textblock}
}

\newcolumntype{P}[1]{>{\centering\arraybackslash}p{#1}}

\newlength{\thickarrayrulewidth}
\begin{document}
	
%
\title{Coarse-to-Fine Registration of Airborne LiDAR Data and Optical Imagery on Urban Scenes}
%
%
%

\author{Thanh~Huy~Nguyen,~\IEEEmembership{Student Member,~IEEE,}
        Sylvie~Daniel,~\IEEEmembership{Senior Member,~IEEE,}
        Didier~Guériot,~\IEEEmembership{Senior Member,~IEEE,}
        Christophe~Sintès,~\IEEEmembership{Senior Member,~IEEE}
        and~Jean-Marc~Le~Caillec,~\IEEEmembership{Senior Member,~IEEE}
\thanks{This work was supported by the Natural Sciences and Engineering Research Council of Canada and the Brittany region, France. \textit{(Corresponding author: Thanh Huy Nguyen.)}}
\thanks{T. H. Nguyen is with the Department of Geomatics, Université Laval, Québec City, QC G1V 0A6, Canada, and also with IMT Atlantique, Lab-STICC, UMR CNRS 6285, F-29238 Brest, France (e-mail: thanh.nguyen@imt-atlantique.fr).}
\thanks{S. Daniel is with the Department of Geomatics, Université Laval, Québec City, QC G1V 0A6, Canada (e-mail: sylvie.daniel@scg.ulaval.ca).}
\thanks{D. Guériot, C. Sintès and J.-M. Le Caillec  are with IMT Atlantique, Lab-STICC, UMR CNRS 6285, F-29238 Brest, France  (e-mail: didier.gueriot@imt-atlantique.fr; christophe.sintes@imt-atlantique.fr; jm.lecaillec@imt-atlantique.fr).}
}

\def\scaleone{0.92}
\def\scaletwo{0.87}

%
%

\markboth{IEEE JOURNAL OF SELECTED TOPICS IN APPLIED EARTH OBSERVATIONS AND REMOTE SENSING,~Vol.~xx, No.~x, MM~yyyy}%
{Shell \MakeLowercase{\textit{et al.}}: Bare Demo of IEEEtran.cls for IEEE Journals}
%


\copyrightstatement


\maketitle

\begin{abstract}
	Applications based on synergistic integration of optical imagery and LiDAR data are receiving a growing interest from the remote sensing community. 
	However, a misaligned integration of these datasets fails to fully profit from the potential of both sensors.
	An optimum fusion of optical imagery and LiDAR data requires an accurate registration. 
	This is a complex problem since a versatile solution is still missing, especially when data are collected at different times, from different platforms, under different acquisition configurations. 
	This paper presents a coarse-to-fine registration method of optical imagery with airborne LiDAR data acquired in such context. 
	Firstly, a coarse registration involves processes of extraction and matching of building candidates from the two datasets.
	Then, a Mutual Information-based fine registration is carried out. 
	It involves a super-resolution approach applied to LiDAR data to generate images with the same resolution as the optical image, and a local approach of transformation model estimation. 
	The proposed method succeeds at overcoming the challenges associated with this difficult context.
	For instance, considering the experimented airborne LiDAR (2011) and orthorectified aerial imagery (2016) datasets, their spatial shift is reduced by 48.15\% after the proposed coarse registration. 
	Moreover, the incompatibility of size and spatial resolution is well addressed by the super-resolution. 
	Finally, a high accuracy of dataset alignment is also achieved, highlighted by a 40-cm error based on a check-point assessment and a 64-cm error based on a check-pair-line assessment.
	These promising results enable further researches for a complete fusion methodology between these datasets in this challenging context.
\end{abstract}

%
\IEEEpeerreviewmaketitle


\section{Introduction}

\label{sec:intro}
\IEEEPARstart{T}{he} 
perception of an environment on the Earth's surface and follow-up exploitations require using multiple sensors to capture specific and  complementary characteristics of this environment \cite{dalla2015challenges}. 
In many areas of remote sensing, observations from heterogeneous sources are coupled and jointly analyzed to achieve a richer description of a scene. 
This approach allows to mutually benefit from their strengths, as well as reducing the data uncertainty and incompleteness relating to each sensor \cite{baltsavias1999comparison,awrangjeb2010automatic,nguyen2017heterogeneous}. 
As a matter of fact, the fusion of multi-source data has become one of the mainstream research topics in the remote sensing community nowadays \cite{dalla2015challenges,ghamisi2019multisource}. 

Light Detection And Ranging (LiDAR) and photogrammetry systems are major sources for fast and reliable spatial data acquisition. 
They provide data that are complementary to each other while the two systems differ fundamentally in their operation and data collection principles. 
The first one is an active sensor while the second is passive.
On the one hand, airborne LiDAR  systems are widely used for providing accurate three-dimensional (3-D) surface information and 3-D geometry of objects and ground elements, in the modality of scattered point clouds (recorded according to range detection principle). 
On the other hand, aerial and satellite photogrammetry supplies rich semantic and texture information, in the form of multispectral images. 
By integrating the two technologies, many applications have been enabled such as building extraction \cite{awrangjeb2010automatic,nguyen2019unsupervised2}, city digital twin construction \cite{Mastin2009}, land use and land cover classification \cite{khodadadzadeh2015fusion} and so on (\cite{Lucas2008,longbotham2012multi}).

\subsection{Motivation}
Over the years, existing works in the domain of data fusion between optical imagery and airborne LiDAR data have addressed dedicated acquisition contexts, in which the respective image and the LiDAR point cloud are already registered and/or they are acquired from the same platform at identical or very close dates. 
For instance, solutions submitted to the 2013 Data Fusion Contest of the IEEE Geoscience and Remote Sensing Society (GRSS) \cite{debes2014hyperspectral} focused on the fusion between LiDAR data and hyperspectral imagery with the same spatial resolution, acquired on two consecutive days. 
The same contest in 2015 \cite{campos2016processing,vo2016processing} involved extremely high resolution LiDAR data and RGB imagery collected from the same aircraft with the sensors being rigidly fixed to the same platform. 
In other words, the solutions submitted to these contests, as well as many others \cite{brell2016improving, Parmehr2014, peng2019automatic}, have not intended to cope with the inherent obstacles of the context where datasets are collected from different platforms with different acquisition configuration (i.e. different flying track, height, orientation, and so on) at different moments and even in different seasons, with different spatial resolutions and levels of detail. 

This research work aims to propose a relevant registration method in this unresolved context.
Table \ref{tab:specs} summarizes the specifications of the sensors and their platforms considered in this work. 
The need for a relevant registration in such a crucial context is exemplified in the work undertaken by Cura \textit{et al.} \cite{cura2017scalable}. 
It also relates to the rise of the availability of data captured by different heterogeneous sensors that requires an efficient integration \cite{ghamisi2019multisource}. 
However, a solution that is versatile enough to overcome this difficult context still remains an unsolved research problem \cite{zhang2017advances}.

\subsection{Challenges}\label{ssec:challenges}
The development of a relevant registration approach in this unresolved context faces many challenges.
\subsubsection{Spatial shift between datasets}\label{sssec:shift}
The first challenge relates to the differences between the dataset point of view and field of view, 
which lead to a significant spatial shift between them. For instance, a spatial shift exists approximately 1-2 meters between the orthorectified airborne image (2016) and the LiDAR data (2011), or up to 40 meters between the Pléiades image and the LiDAR data (2011). 
According to our literature review, a coarse registration, which is necessary to reposition the two datasets, has not been rigorously studied by existing works. 
This step is often inadvertently bypassed using the dataset geospatial coordinates provided by a GPS/IMU system \cite{Mastin2009,parmehr2013automatic,peng2019automatic}.

\subsubsection{Uncertainty, imprecision and incompleteness}
Distortions in the information extracted from optical images can be caused by radiometric errors like sensor sensibility, illumination changes, atmospheric effects, and geometric errors such as relief displacement, occlusions or shadows \cite{abayowa2015automatic}. 
On the other hand, points may be missing in the LiDAR data due to occlusion or presence of water \cite{elberink2008problems}. 
These errors, distortions and missing data from each of the two datasets induce incompleteness, imprecisions and uncertainties within the registration and fusion processes of these data \cite{lahat2015multimodal}.

\subsubsection{Spatial resolution and level of detail}\label{sssec:spatial resolution}
There are significant differences in spatial resolution and level of detail between the airborne or satellite imagery and LiDAR data. 
For example, as highlighted in Table \ref{tab:specs}, the considered LiDAR datasets in 2011 and 2017, respectively, have a point spacing 70 cm and 35.4 cm. 
On the other hand, the aerial image (2016) has a ground sampling distance (GSD) of 15 cm, whereas that of the Pléiades panchromatic (PAN) and multispectral (MS) images are 50 cm and 2 meters, respectively.
They provoke the scene elements and objects into appearing differently on different datasets. 
Such differences affect the appearance of the same scene elements to be different on the two datasets, making it difficult to determine and extract the corresponding features between them \cite{castorena2016autocalibration}.
This issue has not always been addressed by existing multi-source registration works.
For instance, the authors of \cite{ye2019fast} and \cite{ye2017robust} proposed a registration framework between LiDAR image and optical image in a context where the datasets always have  the same spatial resolution.
Among the proposed solutions to overcome the spatial resolution and level of detail differences between the datasets, some involve a multi-resolution approach \cite{Bunting2010,Lucas2008} or a resampling step \cite{abayowa2015automatic,uss2016multimodal}.

\begin{table*}[t]
	\centering
	\caption{Sensor and platform specifications. The boldface rows highlight the differences between datasets notably concentrated in this paper.}\label{tab:specs}
	\renewcommand{\arraystretch}{1.25}
	\scalebox{\scaletwo}{%
		\sffamily
		\begin{tabular}{p{0.25\linewidth}|c|c|c|c}
			\hline
			& \multicolumn{2}{c|}{Optical imagery} & \multicolumn{2}{c}{LiDAR data} \\
			\hline
			& Aerial image (2016) & Satellite image (2015) & Airborne LiDAR (2017) & Airborne LiDAR (2011) \\
			\hline
			Principle & Passive & Passive & Active& Active \\
			\hline
			Device (Camera/LiDAR) & Vexcel UltraCAM Xp & \multirow{2}{*}{Pléiades HR} & Optech ALTM Galaxy & Optech ALTM Gemini \\
			- Platform & - Piper Navajo  & & - Piper Aztec & - Piper Navajo \\
			\hline
			Sensor design & Time-Delay Integration Camera & Pushbroom & Whiskbroom & Whiskbroom \\
			\hline
			\textbf{Acquisition dates (season)} & \textbf{June 2016 (summer)}  & \textbf{June 2015 (summer)} & \textbf{May-June 2017 (summer)} & \textbf{Oct.-Nov. 2011 (winter)} \\
			\hline
			\textbf{Flying height} & \textbf{2955 m} & \textbf{695 km} & \textbf{1300 m} & \textbf{950 m}   \\
			\hline
			\multirow{2}{*}{\textbf{Swath width}} &  \textbf{2597 m (cross-track)} & \multirow{4}{*}{\textbf{20 km}} & \multirow{2}{*}{\textbf{946 m}} & \multirow{2}{*}{\textbf{620 m}} \\
			&  \textbf{1697 m (along-track)}  & &  & \\
			\cline{1-2}\cline{4-5}	
			\multirow{2}{*}{\textbf{Field-Of-View (FOV)}} & \textbf{55$^\circ $ (cross-track)} & & \multirow{2}{*}{\textbf{20$^\circ$}} & \multirow{2}{*}{\textbf{20$^\circ$}}  \\
			&  \textbf{37$^\circ $ (along-track)} & & & \\
			\hline
			Instantaneous & 0.17 mrad (cross-track) & 1  \textmu rad (PAN) & \multirow{2}{*}{-} & \multirow{2}{*}{-} \\
			Field-Of-View (IFOV) &  0.17 mrad (along-track) & 4 \textmu rad (MS) & & \\
			\hline
			Laser repetition rate & -  & -  & 350 kHz & 100 kHz \\
			\hline
			Scan frequency & - & - &  70 Hz & 50 Hz  \\
			\hline
			Laser beam divergence & - & - & 0.25 mrad & 0.25 \& 0.8 mrad \\
			\hline
			Spectral bands/ & \multirow{2}{*}{R, G, B, NIR} & 470-830 nm (PAN) & \multirow{2}{*}{1064 nm} & \multirow{2}{*}{1064 nm} \\ 
			Laser wavelength & &  R, G, B, NIR (MS) & & \\
			\hline
			Number of returns per pulse  & -&-& 4 & 4  \\
			\hline
			Point classification & - & - & U, G, LV, MV, HV, B\footnotemark & U, G, LV, MV, HV \\
			\hline
			\textbf{Point spacing (point density)} & - & - & \textbf{35.4 cm (8 points/m$^2$)} & \textbf{70 cm (2 points/m$^2$)}\\
			\hline
			\textbf{Ground sample distance} & \textbf{15 cm} & \textbf{50 cm (PAN), 2 m (MS)} &-& - \\
			\hline
			Laser beam footprint size & - & - & 33 cm & 23 cm \\
			\hline
			Lateral overlapping & 35\% & - & 65 \% & 30\% \\
			\hline
			Longitudinal overlapping & 60\% & - & - & - \\
			\hline
			\multirow{3}{\linewidth}{Theoretical horizontal accuracy (RMSE)} & \multirow{3}{0.25\linewidth}{\centering 13-16.5 cm (at perspective centers), 4-5.5 cm (at control points) after Aerotriangulation} & \multirow{3}{3.25cm}{1 m (with ground control points) and 3 m (without ground control points)}
			& \multirow{3}{2.4cm}{\centering 1/7500$ \times $Altitude, i.e. $ \approx $17 cm} 
			&  \multirow{3}{2.4cm}{\centering 1/5500$ \times $Altitude, i.e. $ \approx $17 cm}\\
			& & &  & \\
			& & &  &  \\
			\hline
			\multirow{2}{\linewidth}{Theoretical elevation accuracy (RMSE)} &  \multirow{2}{*}{-} &  \multirow{2}{*}{-} &  \multirow{2}{*}{3-20 cm} &  \multirow{2}{*}{5-35 cm}\\
			&&&&\\
			\hline
	\end{tabular}
}

	\footnotesize{$ ^1 $Classification of LiDAR point cloud: unclassified (U), ground (G), low vegetation (LV), medium vegetation (MV), high vegetation (HV) and building (B).}
\end{table*}

\subsubsection{Relevance of registration features}\label{sssec:choice_feature}
The nature of a scene, either in urban or natural environment, conditions strongly the entities within the datasets that would be relevant to perform the registration \cite{Mishra2012}. 

\subsubsection{Accuracy of dataset registration}\label{sssec:accuracy}
When performing the fusion of airborne LiDAR data and optical imagery, even a small misalignment between them can lead to an unfavorable impact on the quality of the integrated product, or a significant reduction of data information content \cite{asner2012carnegie}. 
Thus, an accuracy level of 1-pixel is recommended for the data set registration \cite{Parmehr2014}. 
As a matter of fact, a \textit{sub-pixel level} of accuracy, assessed by measuring the distances between control points, is usually preferred for a  \textit{good} registration.
However, such a qualitative criterion is difficult to achieve because the image pixel resolution can vary from several dozens of centimeters to several meters depending on the platform (i.e. airborne versus satellite). 
Current works in the literature involve resulting discrepancies between the registered datasets ranging from 45 to 50 cm \cite{brell2016improving,liu2016linear}. They state that such discrepancies are a decent and desirable registration accuracy.

\subsection{Contribution}
This paper addresses the need for a {versatile} and relevant registration approach able to overcome the aforementioned challenges. 
The versatility of our proposed method is reflected through its capability of registering the datasets that are not acquired simultaneously, nor from the same platform and same acquisition configuration, nor having same spatial resolution. 
These assumptions are crucial to the existing works \cite{brell2016improving, Parmehr2014, peng2019automatic, ye2017robust, ye2019fast}.
It should be noted that the proposed method does not aim to address every scene possible, as we focus on a registration on urban scenes.
In this regard, we propose a coarse-to-fine registration approach. 
\begin{itemize}
		\item Firstly, a coarse registration is performed to reposition the datasets closer to each other.
		It addresses the challenge of spatial shifts between datasets which is problematic but usually overlooked \cite{Mastin2009,parmehr2013automatic,peng2019automatic,liu2016linear}. 
		In this paper, we present a coarse registration relying on the primitives that are buildings. 
		
		\item Secondly, a fine registration is carried out based on a local transformation model estimation.
		It is enabled by a super-resolution approach applied to LiDAR data in order to generate images with the same resolution as the optical image. This approach is devoted to overcome the hindering caused by the spatial resolution difference between datasets.
\end{itemize}

Such a coarse-to-fine approach is necessary in order to register an airborne LiDAR dataset with an optical image. 
The mentioned coarse registration aims to reposition the two datasets in a fast but reliable manner. 
As a result, a global transformation model, composed of a set of coarsely estimated camera pose parameters, is determined. 
Even though the global transformation does not permit the dataset to be precisely registered, it narrows down the search space for optimal camera pose parameters from an initial set of values during the fine registration.  
However, the main drawback of this feature-based coarse registration is that the building primitives are not distributed evenly throughout the datasets. Hence, the global transformation has the tendency to prioritize a region exhibiting more primitives than others.
Therefore, we propose a subsequent fine registration that focuses on determining the optimal parameters for each local region of the considered urban area. 
Such a local approach brings two benefits, namely a higher registration accuracy and a reduced computational cost of this fine registration. 
Then, we also propose a refinement of locally optimized transformation models, in order to avoid conflicts between them.
Lastly, the proposed method relies on tailored series of well-known processes and algorithms while avoiding complicated and labor-intensive processes.

The remainder of this paper is structured as follows:  
a brief review of existing works related to the registration of optical imagery and airborne LiDAR data is provided in Section \ref{sec:related-work}. Then, Section \ref{sec:methodo} presents the proposed methodological approach, consisting of two parts: coarse registration, then fine registration. Then, multiple quantitative assessments involving different datasets are presented and discussed in Section \ref{sec:results}. Finally, Section \ref{sec:conclusions} provides conclusions and perspectives of this work.

\section{Literature Review}\label{sec:related-work}
Accurate registration of LiDAR data and optical imagery is the crucial prerequisite to any data fusion applications using them \cite{brell2016improving}. 
The majority of automatic methods for registering such datasets can be classified into two categories, namely area-based and feature-based methods. 
On the one hand, area-based methods determine the optimal pose of the camera by maximizing a statistical similarity, e.g. Mutual Information (MI), between the values of optical image pixels and LiDAR-derived image pixels \cite{Parmehr2014,Mastin2009,Bunting2010}. 
The LiDAR-derived image is either a Digital Surface Model (DSM), an intensity image, or an image of \textit{pdet}  (probability of detection) attributes derived from the LiDAR point cloud \cite{Mastin2009}.
Their main drawbacks, in addition to the high computational cost, are the necessities for the datasets to be spatially close to each other, as well as to have the same resolution and display similar intensity characteristics. 
For instance, the similarity of characteristics between two-dimensional (2-D) images and normals to a 3-D surface has been shown to be of paramount importance for area-based registration methods \cite{viola1997alignment}.

On the other hand, feature-based methods establish correspondence between the datasets based on available distinguishable features. 
They involve feature extraction algorithms and feature matching strategy \cite{Ronnholm2012,Ding2008,Wong2008,palenichka2010automatic}. 
The employed features can be either from built environment, such as corner points, break lines and planar surfaces found in man-made objects, or natural features like trees, bushes and ground surface features. 
In general, features from built environment usually yield higher accuracy result than natural features \cite{Ronnholm2012}.

Wong and Orchard \cite{Wong2008} proposed a registration method between LiDAR data and optical image, assuming that they are two images of the same resolution. From the LiDAR data, it is an image of laser return intensity data.
This method consists in using a modified Harris corner detector to extract control points from the two images. 
Then, a Fast Fourier Transform-accelerated exhaustive search for correspondences among all extracted control points is carried out. 
However, this method fails to produce accurate registration result in the case of very high resolution images \cite{Mishra2012}.
Palenichka and Zaremba  \cite{palenichka2010automatic} proposed a registration method between LiDAR-derived DSM and optical imagery. 
It involves an automatic extraction of salient points from both the DSM and the optical image that allows the discrimination of the objects of interest from the background.
This method facilitates the automatic selection of control points that also works on natural scenes. 
According to \cite{Mishra2012}, the high computational cost and the lack of concern for the relief displacement are the drawbacks of this method. 
Liu \textit{et al.} \cite{liu2016linear} proposed a registration method between airborne LiDAR data and UAV (Unmanned Aerial Vehicle) remote sensing imagery, based on 3-D and 2-D line segments extracted, respectively, from the LiDAR point cloud and the image.
For each 3-D line segment, a number of 2-D line segments are extracted from the same location on the image.
Then, a manual selection is carried out to yield the correspondences (i.e. the conjugated line segments). 
Such a manual approach is prone to human bias. 
Also, this method does not account for the potential spatial shift between the datasets. 
Therefore, it could work on the datasets with a small spatial shift, but fails for large spatial shifts.

Many studies have proposed to use different features to increase the registration accuracy. 
For example, Huang \textit{et al.} \cite{Huang2015} proposed a registration method using two different features at two scales, i.e. a line network of roads extracted using  $k$-means clustering at the first scale, and building corners at the finer scale. 
However, the use of $k$-means clustering as an unsupervised classification on aerial images is seemingly too simple  to extract roads effectively. 
Ding \textit{et al.} \cite{Ding2008} performed a coarse-to-fine approach to register oblique aerial image and LiDAR data based on vanishing points estimated from parallel vertical building edges at the coarse level, and then based on building corners at the fine level. 
While the vertical vanishing points can be estimated using oblique images, this can hardly be done using vertical aerial and satellite images, as well as orthorectified images. 
A similar coarse-to-fine approach is also proposed by the authors of \cite{brell2016improving} to register hyperspectral image and LiDAR data simultaneously acquired from the same aircraft. 
First, Scale-Invariant Feature Transform (SIFT) \cite{lowe2004distinctive} keypoint detector is used to determine tie points between the LiDAR data and the hyperspectral image.
Then, an area-based optimization is carried out to find optimal camera pose parameters. 
Within a small range from the values coarsely estimated using the tie points, these parameters are then refined based on the minimization of a cost function.
Such cost function is the zero-mean sum squared distances calculated between the pixels of the hyperspectral image and the image generated from LiDAR intensity data using a ray-tracing module. 
However, this method does not address the registration between the datasets acquired separately, in which the spatial shift between the two datasets can be problematic to the tie-point-based registration.
Also, there is a potential issue due to the spatial resolution of the LiDAR data for generating a suitable image for the area-based optimization.
This issue was not addressed in their work.

In conclusion, all the methods reviewed in this section either assume that the airborne LiDAR data and the optical imagery data are spatially close to each other, have been recorded simultaneously (or on very close dates), and/or have similar spatial resolution and level of detail. 
These constraints have been previously discussed (see \ref{ssec:challenges}) to be challenging to a registration method in the considered context.
To the best of our knowledge, a method explicitly devoted to the registration of LiDAR and image datasets acquired from two different platforms, with different configurations at different times and even seasons has not yet been proposed. 
In what follows, we present how our method is able to achieve such purposes.

\section{Proposed Method}\label{sec:methodo} 

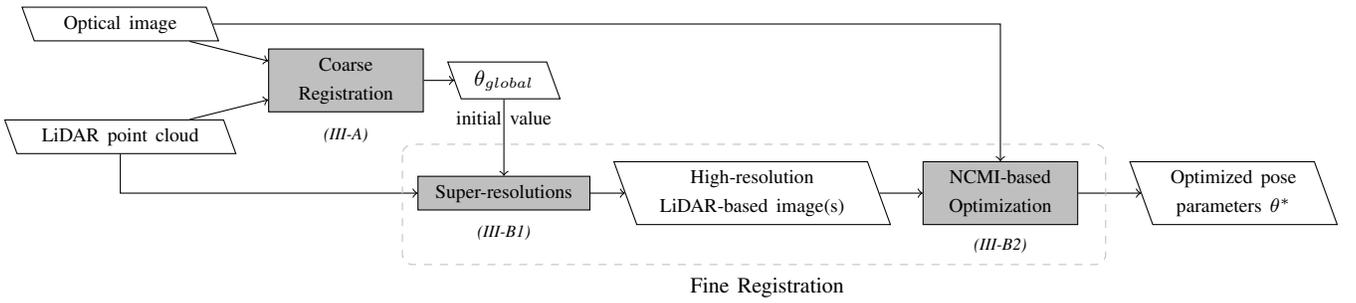
\begin{figure*}[t]
	\centering
	\begin{tikzpicture}[every text node part/.style={align=center},every node/.style={scale=0.92}]

	\node[trapezium,draw,trapezium stretches=false,trapezium left angle=110, trapezium right angle=70,minimum height=0.35cm,text width=2.75cm] (Input1) at (-3,-1.5) {\footnotesize LiDAR point cloud};
	
	\node[trapezium,draw,trapezium stretches=false,trapezium left angle=110, trapezium right angle=70,minimum height=0.35cm,text width=2.25cm]  (Input2) at (-3,0) {\footnotesize Optical image};
	
	\node[draw,fill=gray,text=black,minimum height=0.7cm,text width=2cm] (CR) at (0,-0.75) {\footnotesize Coarse Registration};
	\node[text=black,minimum height=0.35cm,text width=1cm] (CR ref) at (0,-1.475) {\scriptsize \textit{(\ref{subsec:cr})}};
	\node[trapezium,draw,trapezium stretches=false,trapezium left angle=110, trapezium right angle=70,minimum height=0.35cm,text width=1cm]
	(Pglobal) at (2.1, -0.75) {\small $ \theta_{global} $};

	\node[draw,fill=gray,text=black,minimum height=0.35cm,text width=2.25cm] (SR) at (2.1,-2.25) {\footnotesize Super-resolutions};
	\node[text=black,minimum height=0.35cm,text width=1cm] (SR ref) at (2.1,-2.77) {\scriptsize \textit{(\ref{sssec:sr})}};
	
	\node[trapezium,draw,trapezium stretches=false,trapezium left angle=110, trapezium right angle=70,minimum height=0.35cm,text width=3.1cm] (Output1) at (5.4,-2.25) {\footnotesize High-resolution LiDAR-based image(s)};
	
	\node[draw,fill=gray,text=black,minimum height=0.35cm,text width=2cm] (Optim) at (8.7,-2.25) {\footnotesize NCMI-based Optimization};
	\node[text=black,minimum height=0.35cm,text width=1cm] (Optim ref) at (8.7,-2.95) {\scriptsize \textit{(\ref{sssec:optim})}};

	\node[trapezium,draw,trapezium stretches=false,trapezium left angle=110, trapezium right angle=70,minimum height=0.35cm,text width=2.1cm]
	(Output) at (11.8,-2.25) {\footnotesize Optimized pose parameters $ \theta^* $};
	
	\draw[->,draw=black] (Input1) -- (CR);
	\draw[->,draw=black] (Input2) -- (CR);
	\draw[->,draw=black] (Input1) |- (SR);
	\draw[->,draw=black] (CR) -- (Pglobal);
	\draw[->,draw=black] (Pglobal) --node[near start,text width=2cm]{\footnotesize initial~value} (SR);
	\draw[->,draw=black]  (SR) -- (Output1);
	\draw[->,draw=black]  (Output1.east) -- (Optim);
	\draw[->,draw=black]  (Input2) -| (Optim);
	\draw[->,draw=black]  (Optim) -- (Output);
	\draw[-,draw=gray,dashed,rounded corners] (0.75, -1.6) rectangle (10.1,-3.2);
	\node[text=black,text width=3cm] (SymbolicApproach) at (5.6,-3.5) {\small Fine Registration};
	\end{tikzpicture}
	\caption{Flowchart of the registration of optical imagery and airborne LiDAR data (NCMI: Normalized Combined Mutual Information). A parenthesis below each procedure block denotes their respective descriptive sub-section.}
	\label{fig:flowchart_all}
\end{figure*}

Fig. \ref{fig:flowchart_all} presents the full flowchart of the proposed method.
Firstly, the coarse registration approach is presented. It aims to reposition the two datasets based on the extraction and matching of building candidates. 
Based on these primitives, a global transformation model is estimated, which is represented by a set of camera pose parameters, denoted by $ \theta_{global} $. 
Secondly, a fine registration based on super-resolution (SR) of LiDAR values and area-based optimization is carried out.
Such SR process takes into account a transformation model (i.e. $ \theta_{global} $ at the first iteration) and generates high-resolution LiDAR-based images. 
Next, a statistical similarity measure, namely Mutual Information (MI) or Normalized Combined Mutual Information (NCMI), between these super-resolved images and the optical image is estimated.
Thus, the estimated MI (or NCMI) value can be considered as a function of the transformation model. 
The maximum value of such measures is expected to be achieved when the involved images (i.e. the optical image and the super-resolved LiDAR-based images) are geometrically aligned \cite{Parmehr2014}. 
As a result, an optimal transformation model associated to this maximum MI (or NCMI) value is determined. 
We describe the two registrations in the two following sub-sections.

\subsection{Coarse Registration}\label{subsec:cr}
\begin{figure}[!t]
	\centering
	\begin{tikzpicture}[every text node part/.style={align=center},every node/.style={scale=0.95}]
	\node[trapezium,draw,trapezium stretches=false,trapezium left angle=110, trapezium right angle=70,minimum height=0.35cm,text width=2.75cm] (Input1) at (0,0) {\footnotesize LiDAR point cloud};
	\node[trapezium,draw,trapezium stretches=false,trapezium left angle=110, trapezium right angle=70,minimum height=0.35cm,text width=2.5cm] (Input2) at (4,0) {\footnotesize Optical image};
	
	\node[draw,fill=gray,text=black,minimum height=0.35cm,text width=2.75cm] (Process1) at (0,-1) {\footnotesize  Building extraction};
	\node[draw,fill=gray,text=black,minimum height=0.35cm,text width=3.25cm] (Process2) at (4,-1) {\footnotesize  Contextualized Meanshift Segmentation};
	
	\node[text=black,minimum height=0.35cm,text width=1cm] (Process1 ref) at (-1.9,-1) {\scriptsize \textit{(\ref{para:be_lidar})}};
	\node[text=black,minimum height=0.35cm,text width=1cm] (Process2 ref) at (6.15,-1) {\scriptsize \textit{(\ref{para:be_opt_img})}};
	
	\node[trapezium,draw,trapezium stretches=false,trapezium left angle=110, trapezium right angle=70,minimum height=0.35cm,text width=2.6cm] (Output1) at (0,-2.25) {\footnotesize Building candidates (3-D regions)};

	\node[trapezium,draw,trapezium stretches=false,trapezium left angle=110, trapezium right angle=70,minimum height=0.35cm,text width=2.6cm]
	(Output2) at (4,-2.25) {\footnotesize Building candidates (2-D segments)};
	
	\node[draw,fill=gray,text=black,minimum height=0.75cm,text width=2.75cm] (Pairing) at (2,-3.5) {\footnotesize Building segment matching};
	\node[text=black,minimum height=0.35cm,text width=1cm] (Pairing ref) at (3.9,-3.5) {\scriptsize \textit{(\ref{para:gtm})}};

	\node[trapezium,draw,trapezium stretches=false,trapezium left angle=110, trapezium right angle=70,minimum height=0.35cm,text width=2.25cm] (Correspondences) at (2,-4.75) {\footnotesize Segment correspondences};
	
	\node[draw,fill=gray,text=black,minimum height=0.35cm,text width=3cm] (TME) at (2,-5.95) {\footnotesize Global transformation model estimation};
	\node[text=black,minimum height=0.35cm,text width=1cm] (Pairing ref) at (4,-5.95) {\scriptsize \textit{(\ref{para:TME})}};

	\node[trapezium,draw,trapezium stretches=false,trapezium left angle=110, trapezium right angle=70,minimum height=0.35cm,text width=1cm] (Params) at (2,-7) {$ \theta_{global} $};
	
	\draw[->,draw=black] (Input1) -- (Process1);
	\draw[->,draw=black] (Input2) -- (Process2);
	\draw[->,draw=black] (Process1) -- (Output1);
	\draw[->,draw=black] (Process2) -- (Output2);
	\draw[->,draw=black] (Output1) -- (Pairing);
	\draw[->,draw=black] (Output2) -- (Pairing);
	\draw[->,draw=black] (Pairing) -- (Correspondences);
	\draw[->,draw=black] (Correspondences) -- (TME);
	\draw[->,draw=black] (TME) -- (Params);
	\end{tikzpicture}
	\caption{Flowchart of the building-based coarse registration between optical image and LiDAR point cloud. A parenthesis next to each procedure block denotes their respective descriptive sub-section.}
	\label{fig:flowchart_CR}
\end{figure}
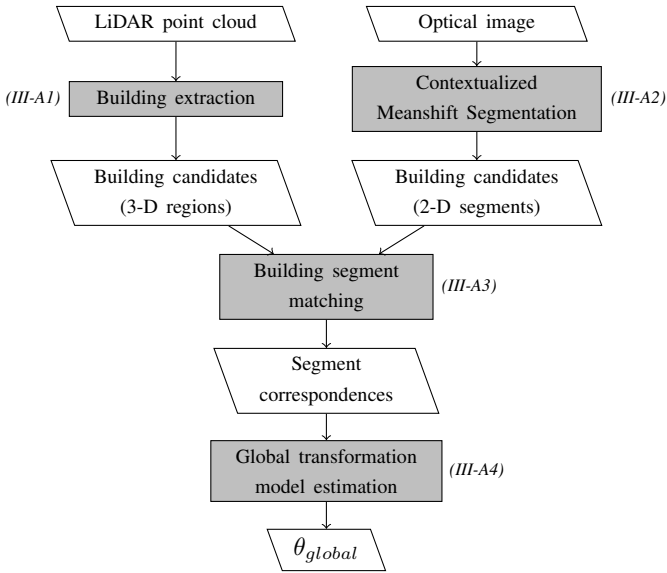

Fig. \ref{fig:flowchart_CR} sums up the proposed coarse registration, which has been originally introduced in our previous work \cite{nguyen2019robust}. 
Man-made structures in urban scenes like buildings are more suitable for accurate registration, compared to natural features \cite{Ronnholm2012}. 
In addition, they remain unchanged through a relatively long period of time (e.g. several years). 
However, in airborne LiDAR datasets, the point density around vertical surfaces like building facades can be low. 
Hence, the localization accuracy of features like building corners and edges is deficient. 
Therefore, our coarse registration method relies on region-based primitives namely buildings.

Different series of processing steps are carried out on the LiDAR and optical image datasets respectively in order to extract buildings. 
On the one hand, we apply a series of processing steps starting with an elevation thresholding on LiDAR point cloud. 
On the other hand, mean shift segmentation \cite{comaniciu2002mean} is performed on the optical image with a contextually chosen bandwidth parameter. 
Further processing is then applied to remove unwanted segments and preserve building-like ones. 
The respective process of building extraction from the LiDAR point cloud and the optical image are described in \ref{para:be_lidar} and \ref{para:be_opt_img}. 
Then, the building candidates from each dataset are matched and yield a set of correspondences (\ref{para:gtm}), which are then used to estimate the global transformation model (\ref{para:TME}). 

\subsubsection{Building  extraction from LiDAR data}\label{para:be_lidar}
\begin{figure}[t]
	\tiny
	\centering
	\begin{subfigure}{0.65\linewidth}
			\centering\includegraphics[trim=0 0 0 0,clip,width=6cm]{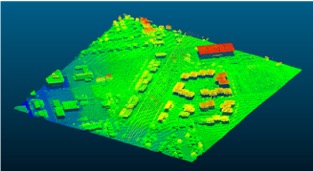}\caption{LiDAR 3D point cloud (visualized by CloudCompare 2.9.1, GPL software)}\label{sfig:3_0}
	\end{subfigure}
	\vspace{0.25cm}
	
	\begin{subfigure}{0.48\linewidth}
		\centering\includegraphics[trim=6.2cm 2.8cm 4.8cm 2cm,clip,height=3.5cm]{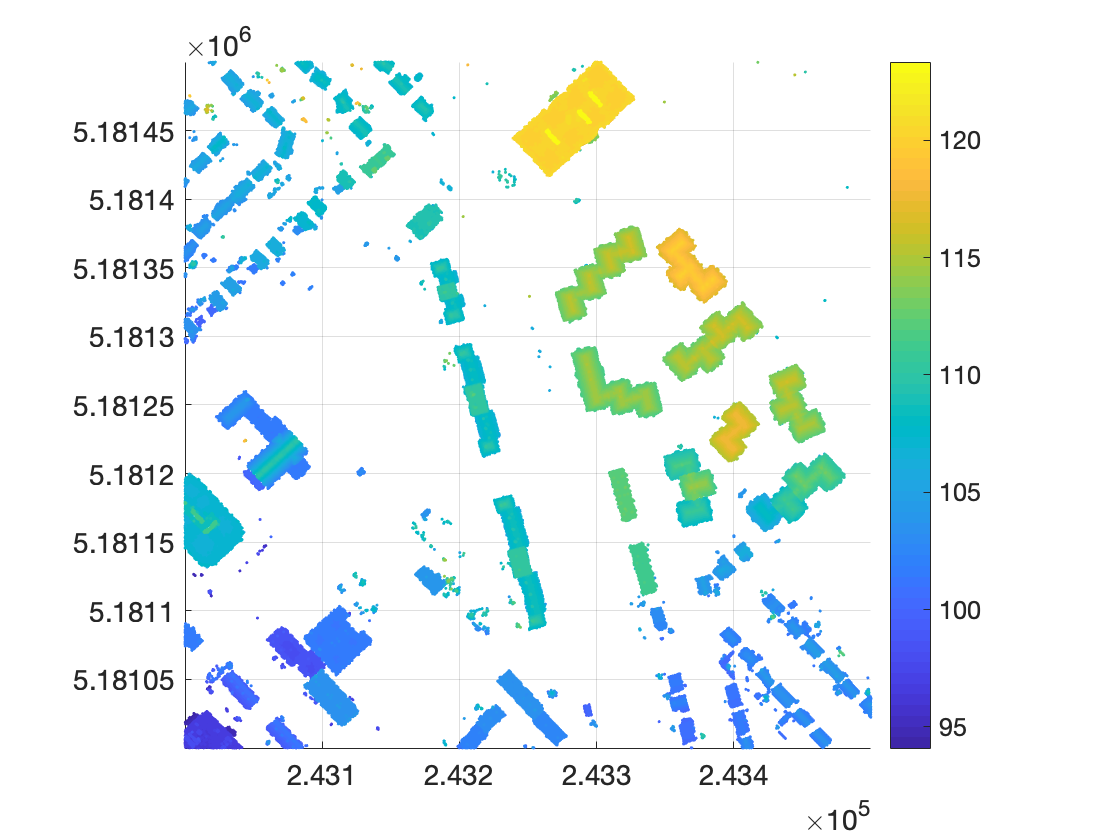}\caption{Grid of non-ground points (color-coded by elevation)}\label{sfig:3_1}
	\end{subfigure}\hspace{0.1cm}
	\begin{subfigure}{0.48\linewidth}
		\centering\includegraphics[trim=4.5cm 2.8cm 4.5cm 2cm,clip,height=3.5cm]{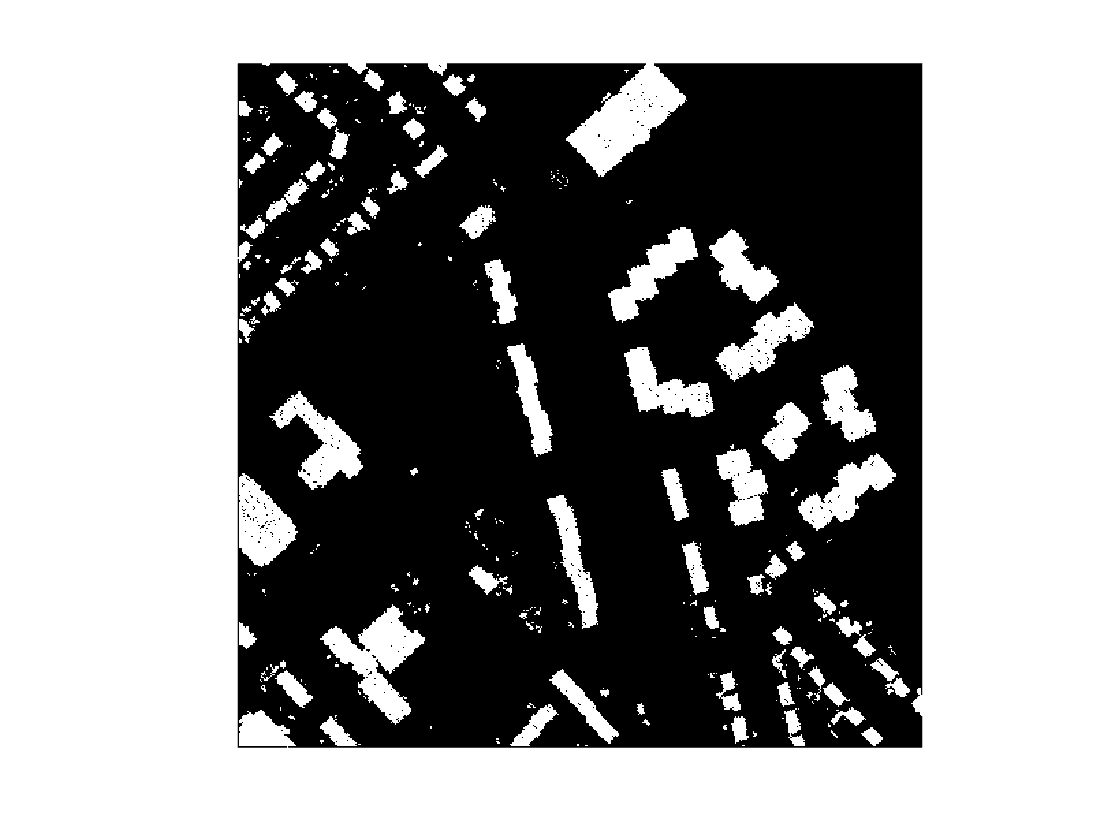}\caption{Binary grid of non-ground points}\label{sfig:3_2}
	\end{subfigure}
	\vspace{0.25cm}

	\begin{subfigure}{0.48\linewidth}
		\centering\includegraphics[trim=4.5cm 2.8cm 6.25cm 2cm,clip,height=3.5cm]{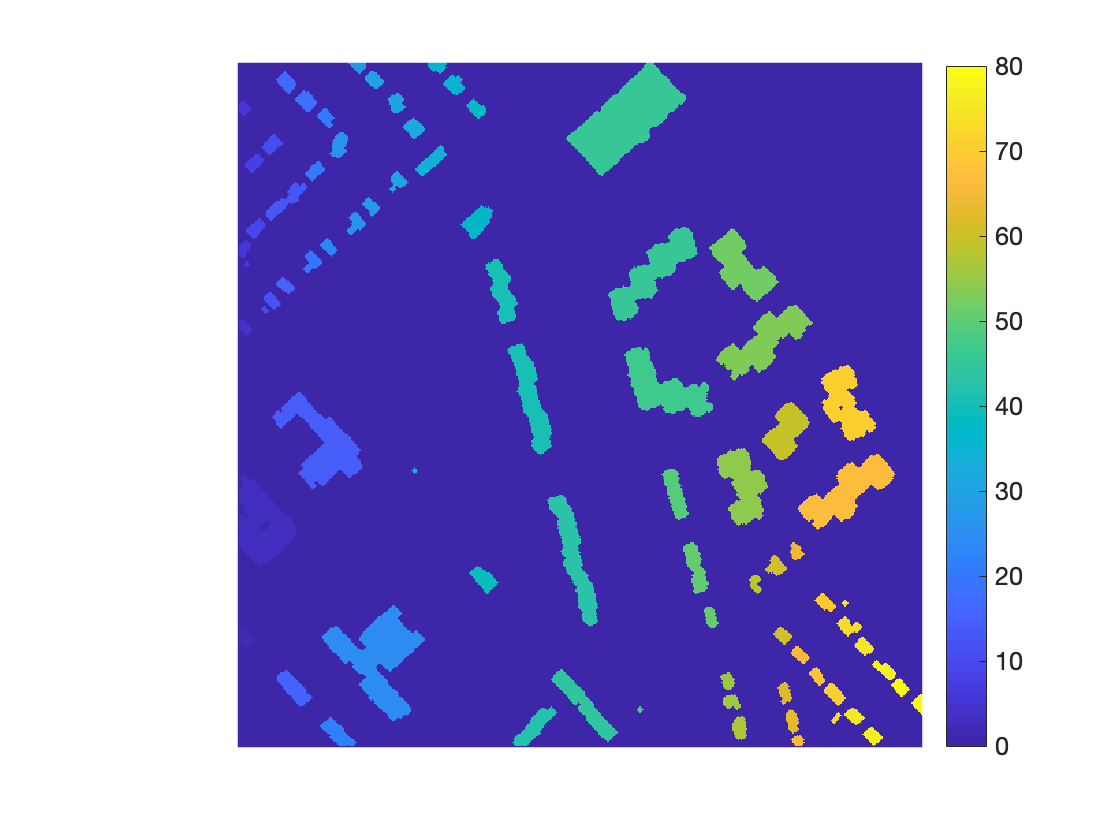}\caption{Labeled segments (distinguished by color)}\label{sfig:3_3}
	\end{subfigure}\hspace{0.1cm}
	\begin{subfigure}{0.48\linewidth}
		\centering\includegraphics[trim=0 0 0 0,clip,height=3.5cm]{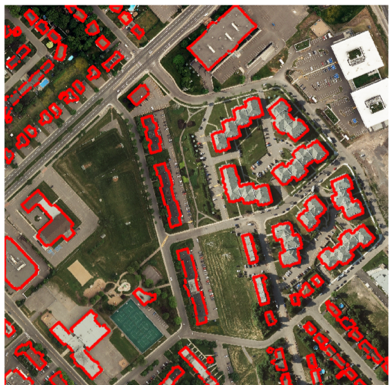}\caption{Extracted building regions (overlapped on optical image)}\label{sfig:3_4}
	\end{subfigure}
	\caption{Illustration of different steps of the building extraction from LiDAR data.}
	\label{fig:seg_lidar}
\end{figure}

The extraction of buildings from LiDAR point cloud is carried out through a series of steps. 
They are depicted in Fig. \ref{fig:seg_lidar}, whereas the input point cloud is shown by Fig. \ref{sfig:3_0}.
First, non-ground points are separated from ground points using an elevation thresholding. 
This thresholding is proposed by many existing works as a necessary initial step \cite{gilani2016automatic}.  
The threshold $ T_e $ is set as follows, $ T_e=H_g + T_{rf}$, where $ H_g $ denotes the ground elevation and $ T_{rf} $ is a relief factor. 
The first value $ H_g $, as proposed by \cite{gilani2016automatic}, can be determined from a Digital Terrain Model (DTM) generated from the LiDAR point cloud data, e.g. by performing \cite{maguya2013adaptive}. 
This DTM generation method allows us to handle complex terrains, such as combination of hills, steep slopes and plateaus.
Also, since the LiDAR point cloud can be classified as described in Table \ref{tab:specs}, e.g. by using \cite{niemeyer2012conditional}, we can measure $ H_g $ by the average elevation of ground points, i.e. $ H_g = \mathrm{mean}(z_g)$ where $z_g$ represents the elevation of ground points. 
The second value $ T_{rf} $ is empirically set to $ T_{rf} = 2.5 $ meters (usual minimum height of a building).

All non-ground points are then vertically projected onto the plane $ z=0 $. 
A raster grid representing these projected points is created (Fig. \ref{sfig:3_1}). 
The resolution of the grid is set according to the LiDAR point cloud density, in order to avoid null-valued pixels. 
For instance, for the LiDAR data 2011 with a point spacing of 70 cm, the resolution of the grid is set to $ 1 $ meter. 
A binary grid of the same resolution is also generated, shown by Fig. \ref{sfig:3_2}. 
Its cell value is set to 1 or 0 according to the presence or absence of projected non-ground points in the cell  (\textquoteleft1\textquoteright: presence, \textquoteleft0\textquoteright: absence).
Then, a morphological opening operator is applied in order to remove small artifacts on the binary grid. 
Remaining grid cells with value set to 1 are grouped into labeled segments based on their connectivity. 
Next, small segments (e.g. smaller than 10 square meters) are removed. 
The resulting grid consists of a number of relatively large labeled segments related to buildings (Fig. \ref{sfig:3_3}). 
These segments are then used to select the building points in the LiDAR point cloud.
A convex hull  is calculated on each set of these 3-D building points, yielding a set of boundary points for each building. 
In Fig. \ref{sfig:3_4}, these building boundaries are shown overlapping on the orthorectified aerial image for visual assessment purpose.

\subsubsection{Building  extraction from optical image}\label{para:be_opt_img}
First, the optical visible image is converted into the CIE L*a*b* color space, since this color space allows a better distinction of objects than RGB color space \cite{hill1997comparative}. 
In this paper, we propose to use mean shift to segment building regions from optical image of an urban area. 
This technique is more efficient than a $ k $-means clustering since the color of building roofs can vary a lot and some roofs have similar color with the surrounding areas or streets.
There exist many other segmentation methods, for example methods based on graph-cut which require different priors, such as connectivity prior \cite{vicente2008graph}, shape prior \cite{freedman2005interactive}, or priors about color of background and foreground pixels given by several brush strokes on an image.
As a matter of fact, the graph-cut-based image segmentation methods require a high amount of user inputs in order to yield accurate results \cite{szeliski2010computer}.
On the other hand, mean shift requires only a value of bandwidth corresponding to the image color range and size of objects to be segmented.
This technique has been extensively used for many years in the field of computer vision and image processing. 
However, its relevance cannot yet be dismissed.
Nevertheless, determining the best bandwidth parameter for mean shift still remains difficult despite many investigated approaches \cite{chacon2013comparison}. 
This parameter can be set adaptively according to the type of urban area (either residential, industrial, mixed, etc.), and the size of objects of interest. 
In other words, a contextualization is needed to set up the mean shift parameter.
Such a contextualization is carried out based on the meaningful information in the observed area, such as an estimated number of buildings and their relative distance---this knowledge is derived from the building extraction process using LiDAR data---as well as the resolution and the color range of the optical image.
Future works will investigate the automation of this step.

\begin{figure}[!t]
	\begin{subfigure}[b]{0.3\linewidth}
		\centering\includegraphics[trim=2cm 2cm 2cm 1cm, clip, height=2.5cm]{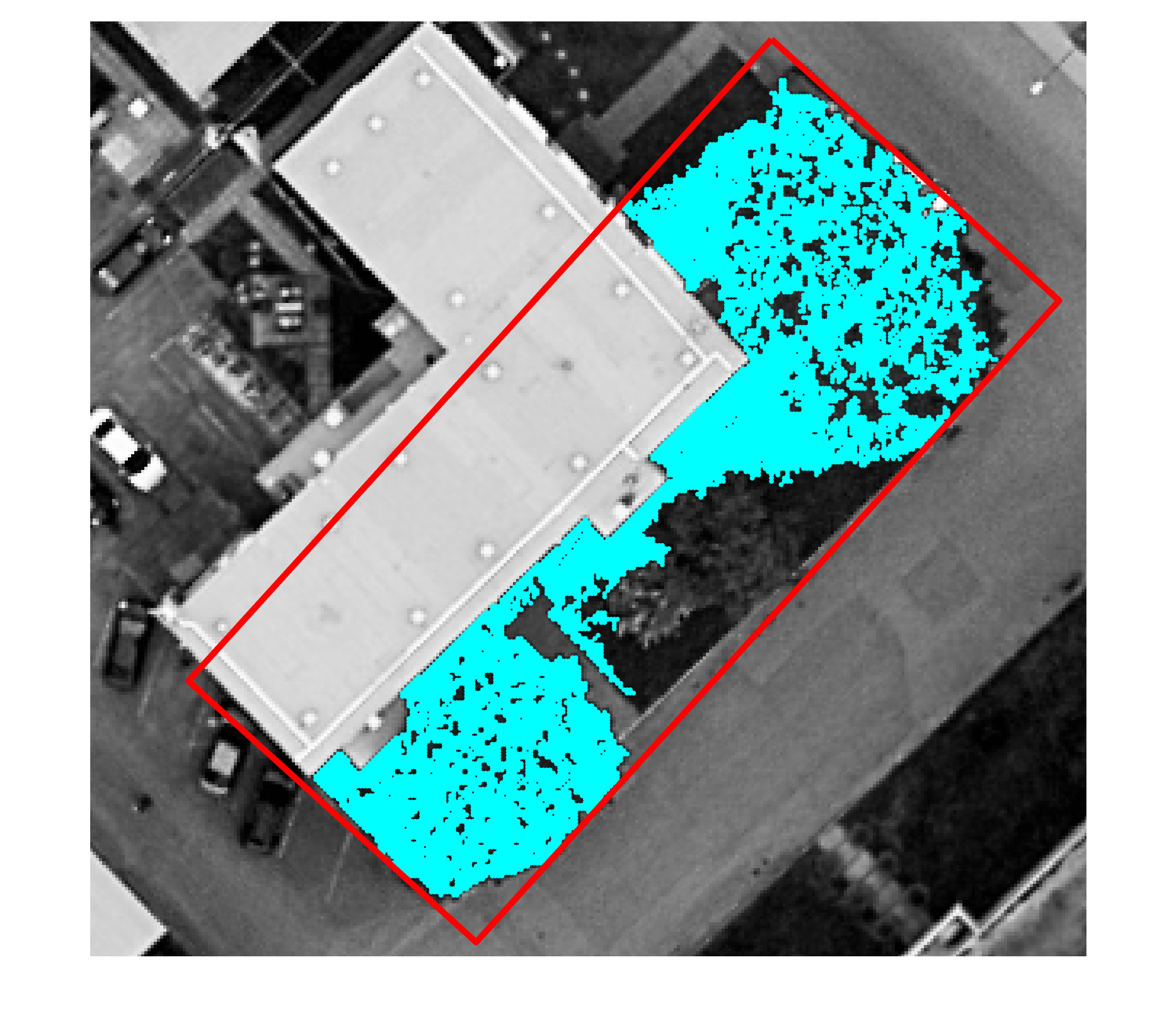}\caption{On a tree segment}
	\end{subfigure}\hspace{0.1cm}
	\begin{subfigure}[b]{0.3\linewidth}
		\centering\includegraphics[trim=2cm 2cm 2cm 1cm, clip, height=2.5cm]{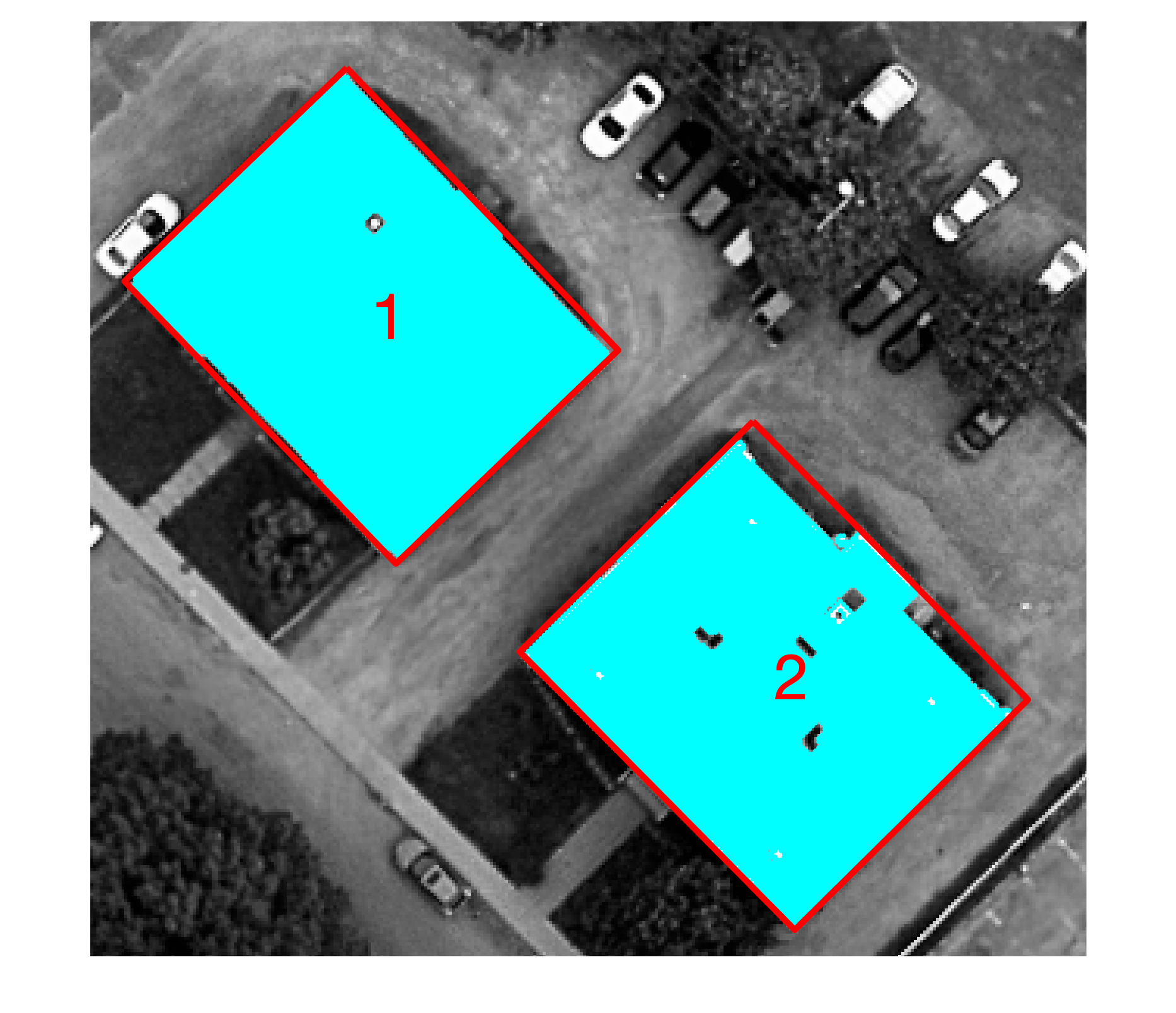}\caption{On building segments}
	\end{subfigure}\hspace{0.1cm}
	\begin{subfigure}[b]{0.3\linewidth}
		\scalebox{0.75}{
		\begin{tabular}{cc}
			\hline
			\textbf{Segment} & \textbf{\%\textsubscript{MBR filling}} \\
			\hline
			Tree & 43.41\%\\
			\hline
			Building 1 &  94.75\%\\
			Building 2 & 91.57\%\\
			\hline
		\end{tabular}}
		\caption{MBR filling percentage}
		\label{tab:percent}
	\end{subfigure}
	\caption{Comparison of the MBR filling percentage between a tree segment versus building segments. The segment pixels are in cyan, whereas the MBR of each segment is in red.}
	\label{fig:MBR_LER}
\end{figure}

Once the mean shift segmentation is performed, a refinement of the extracted segments is carried out. 
First, we compute the size of the segments, and remove the small ones, since they usually correspond to trees and cars. 
Large segments corresponding to street regions are similarly removed. 
This filtering is simple and efficient \cite{gavankar2018automatic}, but depends on the image resolution. 
Therefore, it needs a manual intervention to be set correctly. 
In this paper, we propose to remove segments whose actual area is smaller than 20 square meters or larger than 2,000 square meters, which are not the typical area of buildings.
Second, we identify the Minimal Bounding Rectangle (MBR) \cite{freeman1975determining} of each of the remaining segments and calculate, using Eq. \eqref{eq:MBR}, the  percentage of their area over the area of the MBR.
\begin{equation}\label{eq:MBR}
	\%_{\textnormal{MBR\_filling}}=\dfrac{\textnormal{area(segment)}}{\textnormal{area(MBR)}}\times 100
\end{equation}
This filling percentage aims to eliminate coarsely the irregular segments such as trees and grass, while retaining highly regular shape building segments. 
Fig. \ref{fig:MBR_LER} depicts a comparison between the MBR filling percentages of two building segments and a tree segment, from which a clear margin between the two types of segment can be observed.
In this paper, a threshold of 50\% for the MBR filling percentages is typically applied.
However, on a scene with numerous irregular shape buildings, this threshold can be relaxed.
It is worth noting here that this MBR-based refinement only acts as a {preliminary} filter.
Although it cannot remove every non-building segment, it allows to effectively eliminate coarsely the irregular segments.
Then, these extracted and refined segments, even with a number of potential outliers, will be fed into the graph-based matching step.

\subsubsection{Graph-based matching of extracted segments}\label{para:gtm}
The two sets of building candidates extracted from the LiDAR and optical image datasets are taken into consideration and matched. 
Regarding the optical image, only the segments having a higher percentage than the fixed threshold are considered as stated in the previous point.  
On the other hand, all building regions extracted from the LiDAR point cloud are taken into consideration. 
The comparison and matching of these segments can be difficult due to several issues.
First, several tree and grass segments wrongly extracted as buildings still remain after the MBR-based segment refinement. 
In addition, the datasets can be relatively distant to each other (as mentioned in \ref{sssec:shift}), making a direct matching of segments based on their location is not suitable. 
Therefore, a matching of segments based on their relative position with respect to their neighbors is more relevant than comparing their individual values, such as location, area, shape similarity, and so on.

A common pattern connecting the centers of neighboring building segments representing their relative spatial arrangement on both datasets is determined using the Graph Transformation Matching (GTM) algorithm \cite{aguilar2009robust}. 
GTM is a graph-based point matching algorithm designed for non-rigid registration between images. 
Compared to a conventional method like RANSAC \cite{fischler1981random}, this algorithm performs a better removal of \textit{outliers}, i.e. wrongly paired buildings in this work.

In practice, both GTM and RANSAC require an initial one-to-one matching of segment centers, which can be carried out based on the positions of vertically projected 3-D building region centers onto the plane $ z=0 $ and the centers of 2-D segments extracted by mean shift segmentation. 
In the specific case of  satellite imagery and LiDAR data where the relative shifts are large (i.e. approximately up to 40 meters), this initial matching is guided by a translation vector. 
It is calculated based on the shift of the largest segment in the area. 
The largest segment is determined relying on the segment absolute area value and its relative area value with respect to other segments.

\begin{figure*}[t]
	\centering
	\begin{subfigure}[b]{0.33\linewidth}
		\centering\includegraphics[trim=3cm 1.5cm 3cm 1cm,clip,width=4.75cm]{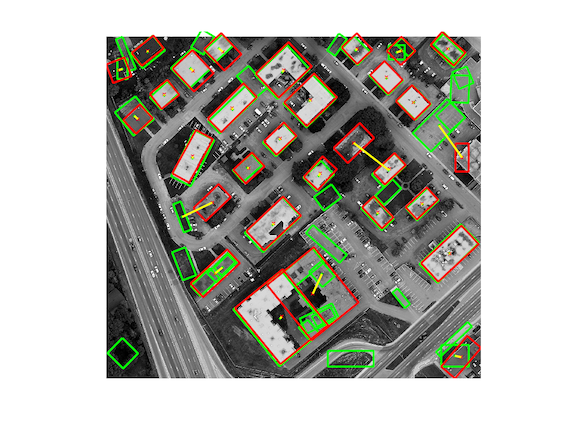}\caption{Initial matching}\label{subfig:initial}
	\end{subfigure}
	\hspace{0.02cm}
	\begin{subfigure}[b]{0.33\linewidth}
		\centering\includegraphics[trim=2.5cm 1cm 1.5cm 0.75cm,clip,width=4.75cm]{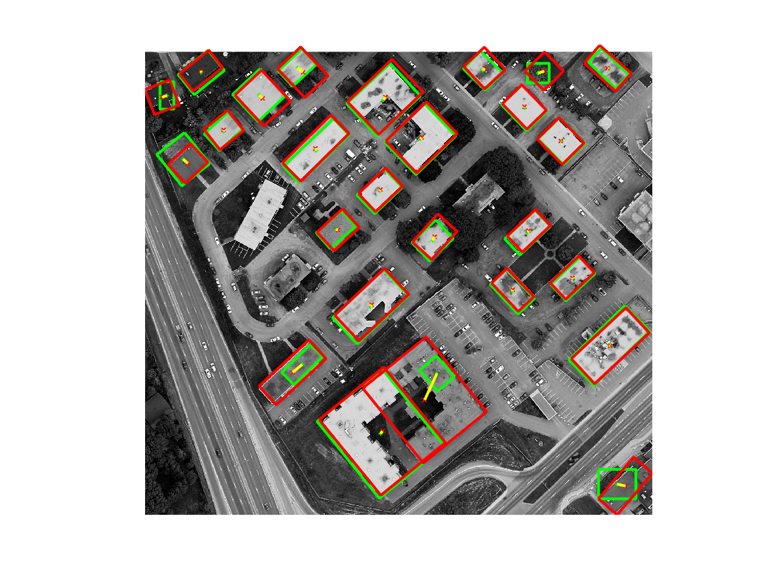}\caption{GTM result}\label{subfig:gtm}
	\end{subfigure}

	\vspace{0.25cm}
	\begin{subfigure}[b]{0.33\linewidth}
		\centering\includegraphics[trim=3cm 1.5cm 3cm 1cm,clip,width=4.75cm]{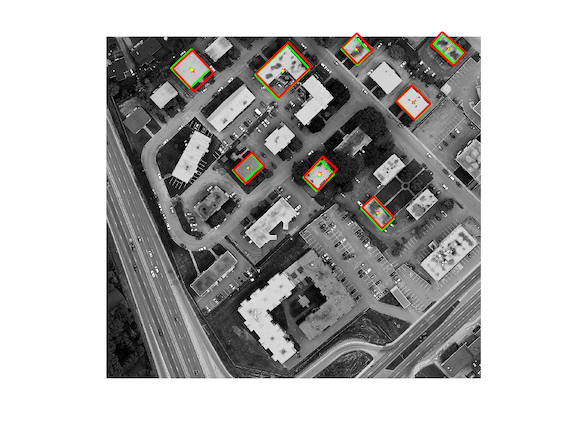}\caption{RANSAC result}\label{subfig:ransac}
	\end{subfigure}
	\hspace{0.02cm}
	\begin{subfigure}[b]{0.33\linewidth}
		\centering\includegraphics[trim=2.5cm 1cm 1.5cm 0.75cm,clip,width=4.75cm]{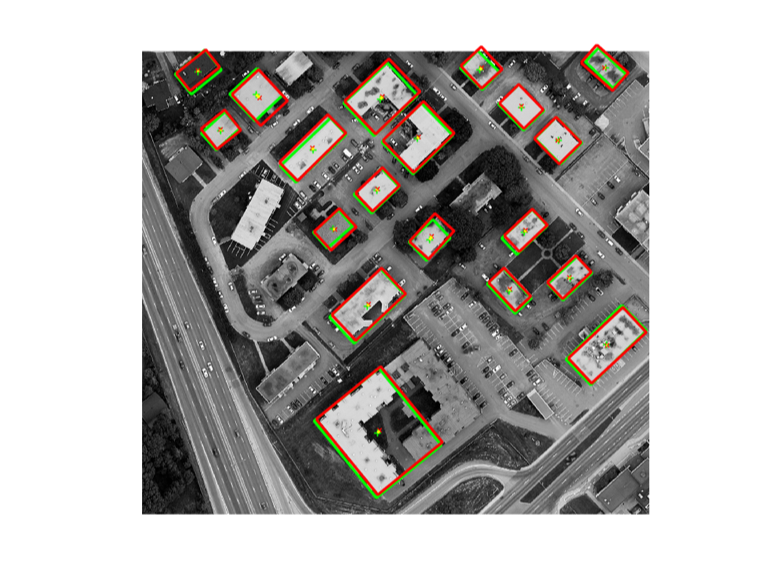}\caption{GTM + Area and direction validation}\label{subfig:gtm_final}
	\end{subfigure}
	\caption{Matching of building segment centers by considering their relative position. The green and red rectangles represent the MBR of the segments extracted from, respectively, optical image and LiDAR point cloud. The yellow lines  connect the centers of the matched segments.}
	\label{fig:GTM}
\end{figure*}

Result of the segment matching is shown in Fig. \ref{fig:GTM}, whereas Fig. \ref{subfig:initial} depicts the initial matching. 
As we could expect, a number of wrongly paired buildings (i.e. {outliers}) result from the initial matching. 
They are originated from the tree and grass segments extracted as buildings, or from the buildings that exist on one dataset but not on the other one.
These outliers are then removed using GTM.
Fig. \ref{subfig:gtm} depicts the result of GTM, whereas Fig. \ref{subfig:ransac} presents the result of RANSAC. 
As we have been considering only the relative position of the segment centers, a refinement of false positives from GTM result is carried out based on the area value and the direction of segments. 
Here, we allow some tolerance for the area value (i.e. a 15\% difference) and direction (i.e. a 2$ ^\circ $ difference)  between paired segments provided by GTM. 
Such tolerance values are chosen empirically. 
Only the pairs of segments having area and direction differences smaller than the tolerances are preserved.
With the selected tolerance values, we consider that only reasonable correspondences of buildings will remain.
The result of this refinement is presented by Fig. \ref{subfig:gtm_final}.

The capability of GTM to cope with high amount of outliers---theoretically up to three times more numerous than the correct pairs \cite{aguilar2009robust}---is advantageous when handling the potential high number of outliers among the extracted segments from the optical image using mean shift.
It is also anticipated to handle well the registration of datasets that were acquired within a large timespan, e.g. several years. 
This temporal variability can lead to significant changes in urban area, such as construction or deconstruction of buildings.

\subsubsection{Global transformation model estimation}\label{para:TME}
Next, the coordinates of the matched building segment centers are used to determine the transformation model between the LiDAR and the imagery datasets. 
It involves in estimating the camera pose internal and external parameters. 
The internal parameters are consisted of the scale factors in the $ x $- and $ y $-coordinate direction, respectively $ \alpha_x  $ and $ \alpha_y  $, the skew parameter $ s $ and the coordinates of the principal point $ (p_x, p_y) $ in terms of pixel dimensions.
On the other hand, the external parameters are the position $(X_0, Y_0, Z_0)$ and the orientation $(\omega, \phi, \kappa)$ of the camera when the image was acquired.
The set of all these parameters of the camera pose is denoted by a vector $ \theta $ henceforth. 

Even though the camera pose  external parameters can be measured by a GPS/IMU system, it is still necessary to reestimate them, since the image can suffer from radiometric and geometric errors as well as undergone an orthorectification process.
Based on the coordinates of the resulting matched building segment centers, a transformation model is estimated using the Gold Standard algorithm for finite projective camera model \cite[p.187]{hartley2003multiple}. 
We denote the set of parameters associated with this estimated global transformation model by $ \theta_{global} $. 

The transformation from 3-D homogeneous coordinates to 2-D homogeneous coordinates is given by the following $ 3\times 4 $ camera matrix,
\begin{equation}\label{eq:P}
		P=KR \left[ I~|~-{C} \right]
\end{equation}
where $ K $ is the camera calibration matrix, $ R $ stands for the rotation matrix describing the orientation of the camera, $ I $ is the identity matrix, and $ C $ denotes the coordinates of the camera center. 
The matrices from Eq. \eqref{eq:P} are defined as follows, 
\begin{subequations}
		\begin{equation}\label{eq:K}
				K=\left[
				\begin{matrix}
				\alpha_x & s & p_x\\
				0 & \alpha_y & p_y\\
				0 & 0 & 1\\
				\end{matrix}
				\right]
		\end{equation}
		\begin{equation}\label{eq:C}
				C=
				\left[
				\begin{matrix}
				X_0 & Y_0 & Z_0\\
				\end{matrix}
				\right]^T
		\end{equation}
		\begin{equation}\label{eq:R}
				R=R_z(\kappa)R_y(\phi)R_x(\omega)
		\end{equation}
		\label{eq:KRC}
\end{subequations}
where $ R_x, R_y, R_z $ are the rotation matrices for rotations around $ x $-, $ y $- and $ z $-axis.
As presented by Eq. \eqref{eq:KRC}, the transformation model involves eleven degrees of freedom, related to the camera internal and external parameters.

\subsection{Fine Registration}\label{subsec:fr}
After coarsely repositioning the datasets, the next step is dedicated to register them precisely. 
An area-based optimization approach is relevant in the present context, in order to determine the optimal set of parameters that enables the most accurate registration \cite{brell2016improving,Parmehr2014}. 
However, this approach involves several constraints, such as the datasets need to be spatially close to each other, as well as to have the same resolution and display similar intensity characteristics. 
As a result of the presented coarse registration and the SR process elaborated in the next sub-section (\ref{sssec:sr}), these constraints are fulfilled.

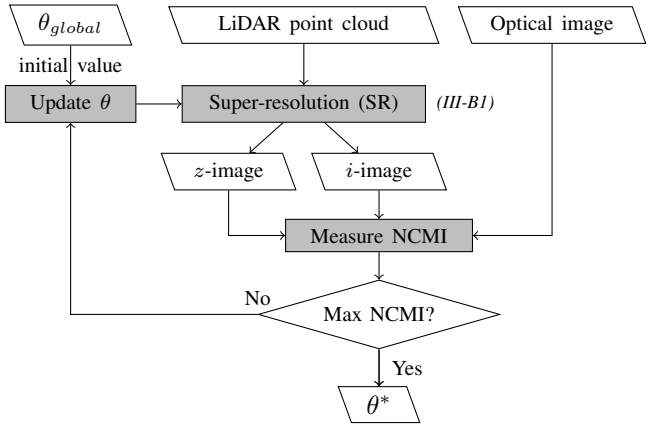
\begin{figure}[t]
	\centering
	\begin{tikzpicture}[every text node part/.style={align=center},every node/.style={scale=1}]
	\node[trapezium,draw,trapezium stretches=false,trapezium left angle=110, trapezium right angle=70,minimum height=0.35cm,text width=3cm] (Input1) at (1.1,0.15) {\footnotesize LiDAR point cloud};
	
	\node[trapezium,draw,trapezium stretches=false,trapezium left angle=110, trapezium right angle=70,minimum height=0.35cm,text width=1.9cm] (Input2) at (4.4,0.15) {\footnotesize Optical image};
	
	\node[trapezium,draw,trapezium stretches=false,trapezium left angle=110, trapezium right angle=70,minimum height=0.3cm,text width=1cm] (Pglobal) at (-2,0.15) {\small $ \theta_{global} $};
	\node[draw,fill=gray,text=black,minimum height=0.35cm,text width=1.5cm] (P) at (-2,-0.9) {\footnotesize Update $ \theta $};
	\node[draw,fill=gray,text=black,minimum height=0.35cm,text width=3cm] (SR) at (1.1,-0.9) {\footnotesize  Super-resolution (SR)};
	\node[text=black,minimum height=0.35cm,text width=1cm] (SR ref) at (3.25,-0.9) {\scriptsize \textit{(\ref{sssec:sr})}};

	\node[trapezium,draw,trapezium stretches=false,trapezium left angle=110, trapezium right angle=70,minimum height=0.35cm,text width=1.2cm](zimg) at (0.1, -1.8) {\footnotesize $ z $-image};
	\node[trapezium,draw,trapezium stretches=false,trapezium left angle=110, trapezium right angle=70,minimum height=0.35cm,text width=1.2cm](iimg) at (2.1, -1.8) {\footnotesize $ i $-image};
	\node[draw,fill=gray,text=black,minimum height=0.35cm,text width=2.25cm] (NCMI) at (2.1,-2.65) {\footnotesize Measure NCMI};
	\node[draw, diamond, aspect=3.5] (NCMI2) at (2.1,-3.7) {\footnotesize Max NCMI?};
	
	\node[trapezium,draw,trapezium stretches=false,trapezium left angle=110, trapezium right angle=70,minimum height=0.35cm,text width=0.5cm] (Result) at (2.1,-4.9) {$ \theta^* $};
	
	\draw[->,draw=black] (Input1) -- (SR);
	\draw[->,draw=black] (Input2) |- (NCMI);
	\draw[->,draw=black] (Pglobal) --node[midway,text width=1.7cm]{\footnotesize initial~value} (P);
	\draw[->,draw=black] (P) to (SR);
	\draw[->,draw=black]  (SR) -- (zimg);
	\draw[->,draw=black]  (SR) -- (iimg);
	\draw[->,draw=black]  (zimg) |- (NCMI);
	\draw[->,draw=black]  (iimg.south) to (NCMI);
	\draw[->,draw=black] (NCMI) -- (NCMI2);
	\draw[->,draw=black] (NCMI2) -- (Result);
	\draw[->,draw=black] (NCMI2) --node[midway,right,text width=0.5cm]{\footnotesize Yes}  (Result);
	\draw[->,draw=black] (NCMI2) -|node[at start,above,text width=1cm]{\footnotesize No}  (P);
	\end{tikzpicture}
	\caption{Flowchart of the proposed fine registration between optical image and LiDAR point cloud. }
	\label{fig:flowchart_FR}
\end{figure}

We propose a fine registration method, summarized by Fig. \ref{fig:flowchart_FR}, which involves a SR applied on the LiDAR data. 
Then, an estimation of local transformation models (sub-section \ref{sssec:optim}) is performed based on the maximization of the NCMI or MI measured between the optical image and the \textit{high-resolution} LiDAR-based images, resulted from the SR. 
The high-resolution term means that these images have the same resolution and size as the optical image. 
NCMI achieves its maximum values when the images are geometrically aligned \cite{Parmehr2014}, yielding an optimal set of camera pose parameters, denoted by $ \theta^* $. 
We describe these points in what follows.

\subsubsection{Super-resolution of LiDAR data}\label{sssec:sr}
LiDAR point cloud is usually significantly subsampled compared to optical image. 
This subsample problem is usually addressed by a sparse reconstruction (e.g. for pansharpening \cite{zhu2013sparse}) or a super-resolution of low-resolution depth maps \cite{lu2015sparse}. 
Thus, we propose a process of transferring and propagating values from LiDAR point cloud onto the frame of the optical image. 
Such process is to generate a rasterized dataset with the same size and spatial resolution as the optical image, thus it is called super-resolution. 
Pixels of the super-resolved image contain the values derived from the LiDAR 3-D points, i.e. altitude values or laser return intensity values. 
The super-resolved image of LiDAR-derived altitude values is called the $ z $-image, whereas the image of intensity values is called $ i $-image.
The purpose of the SR is also to provide an approach to neutralize the sampling density difference between the two datasets, thus facilitate the area-based fine registration of them.

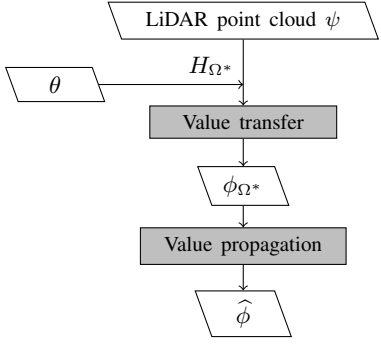
\begin{figure}[t]
	\centering
	\begin{tikzpicture}[every text node part/.style={align=center},every node/.style={scale=1}]
	\node[trapezium,draw,trapezium stretches=false,trapezium left angle=110, trapezium right angle=70,minimum height=0.35cm,text width=3cm] (Input1) at (1,-1.3) {\footnotesize LiDAR point cloud $ \psi $};
	\node[trapezium,draw,trapezium stretches=false,trapezium left angle=110, trapezium right angle=70,minimum height=0.35cm,text width=0.75cm] (P) at (-1.5,-2.15) {\small $ {\theta} $};
	\node[draw,fill=gray,text=black,minimum height=0.35cm,text width=2.25cm] (transf) at (1,-2.65) {\footnotesize  Value transfer};
	\node[trapezium,draw,trapezium stretches=false,trapezium left angle=110, trapezium right angle=70,minimum height=0.35cm,text width=0.6cm] (phi om) at (1,-3.5) {\small $ \phi_{\Omega^*} $};
	\node[draw,fill=gray,text=black,minimum height=0.35cm,text width=2.5cm] (propa) at (1,-4.3) {\footnotesize Value propagation};
	\node[trapezium,draw,trapezium stretches=false,trapezium left angle=110, trapezium right angle=70,minimum height=0.35cm,text width=0.6cm] (phi) at (1,-5.2) {\small $ \widehat{\phi} $};
	\draw[->,draw=black] (Input1) -- (transf);
	\draw[->,draw=black] (transf) -- (phi om);
	\draw[->,draw=black] (phi om) to (propa);
	\draw[->,draw=black]  (propa) -- (phi);
	\draw[->,draw=black] (P) --node[near end,above,text width=0.5cm]{\small $ H_{\Omega^*} $}  (1,-2.15);
	\end{tikzpicture}
	\caption{Overview of the super-resolution process, to generate a high-resolution LiDAR-based image ($ z $- or $ i $-image).}
	\label{fig:flowchart_SR}
\end{figure}

\paragraph{Mathematical notation}
The inputs of the SR process are the LiDAR point cloud, a transformation model, the frame of reference and the size of the optical image. 
We denote the optical image by $ u \in \mathbb{R}^{n_x \times n_y \times 3} $, where $ n_x $ and $ n_y $  are respectively the number of rows and columns. 
The LiDAR point cloud is represented by $ \psi \in \mathbb{R}^{m \times 4} $, where $ m $ is the number of LiDAR points. 
Each point contains three spatial coordinates $ (x,y,z) $ and a laser return intensity value $ i $. 
We also use $ \psi^z \in \mathbb{R}_+^{m}$ to denote the column of altitude values of the LiDAR points; 
whereas $ \psi^i \in \{0, 1,..., 255\}^{m}$ stands for the intensity value of LiDAR points. 
For the sake of simplicity, we use the same notation $ \phi  $ to denote the result of the SR, i.e. the $ z $-image and $ i $-image.  
During the process, $ \phi $ is vectorized into a column vector of $ n={n_x \times n_y} $ elements. 
Fig. \ref{fig:flowchart_SR} describes the principle of the proposed SR.

\paragraph{Transfer of LiDAR values}
First, LiDAR values, i.e. altitude and laser return intensity values, are projected onto the optical image space using the transformation model associated with camera pose parameters $ \theta $. 
At the first iteration of the fine registration, $ \theta $ is given by $ \theta_{global} $ obtained from the coarse registration. 
Mathematically, the value transfer is presented by the following equation,
\begin{equation}\label{eq:val_transfer}
		\phi_{\Omega^*}=H_{\Omega^*}\psi^z \textnormal{ or }  \phi_{\Omega^*}=H_{\Omega^*}\psi^i
\end{equation}
where  $ \Omega^* $ and $ \Omega $ denote, respectively, the subsets containing the indices of pixels from $ \phi $, having or not an associated altitude value (or intensity value) transferred from $ \psi $. 
Thus, $ \phi_{\Omega^*} $ and $ \phi_\Omega $, respectively, denote the sub-vector containing the pixels with and without a transferred altitude value; whereas $ \phi $ denotes the vector containing all pixels. 
The dimension of $ \phi_{\Omega^*} $ and $ \phi_\Omega $, respectively, are $ m\times 1 $ and $ (n-m)\times 1 $.
The matrix $ H_{\Omega^*} $ associated to the camera pose parameters $ \theta $, is an index matrix allowing selecting only the pixels whose values are transferred from the LiDAR point cloud. 
It is computed based on the projection related to $ \theta $ (Eq. \eqref{eq:P}) of the LiDAR 3-D point cloud onto the 2-D optical image space.
Next, the transferred values are propagated to their neighboring pixels.

\paragraph{Propagation of transferred LiDAR values}
The propagation of transferred values is carried out through the minimization of a cost function $ \mathcal{F}(\phi) $,  defined by Eq. \eqref{eq:zvtp1}. 
It is composed of the sum of squared directional gradients (SSDG) of $ \phi $, and a $ \mathcal{L}^1 $-norm term to promote the sparsity of $ \phi $, subjecting to the values transferred from the point cloud (described by Eq. \eqref{eq:val_transfer}). 
\begin{equation}\label{eq:zvtp1}
		\begin{array}{c}
				\widehat{\phi}=  \underset{\phi}{\arg\min}~\mathcal{F}(\phi) \\[10pt]
				\textnormal{with } \mathcal{F}(\phi) = \underbrace{\left\| \nabla_x\phi\right\|^2_{2} + \left\| \nabla_y\phi\right\|^2_{2}}_{\vspace{0.25cm}f_{\mathrm{SSDG}}(\phi)}+\lambda\left\|\phi\right\|_{1},\\[20pt]
				\textnormal{subject to }  \phi_{\Omega^*}=H_{\Omega^*}\psi^z \textnormal{ or }  \phi_{\Omega^*}=H_{\Omega^*}\psi^i 
		\end{array}
\end{equation}
where $ \left\|\cdot\right\|_{p} $ stands for the $ \mathcal{L}^p $-norm, $ \nabla_x $ and $ \nabla_y $ represent the directional gradient operators along the $ x $- and $ y $-axis, whereas the parameter $ \lambda $ controls the amount of the $ \mathcal{L}^1 $-regularization.

Our SR approach is inspired by the work of Castonera \textit{et al.} \cite{castorena2018motion}. 
However, they proposed a cost function that is solely defined by SSDG for the fusion of terrestrial LiDAR data with optical imagery. 
It is based on hypothetical characteristics of a depth map, namely that the magnitude and occurrence of depth discontinuities inside such depth map should be minimum. 
The advantage of using this cost function is its convexity and ease to compute. 
Castonera's method showed good results in propagating depth values across homogeneous regions. 
However, the mentioned hypothetical characteristics are not suitable in an airborne context, where off-terrain objects like buildings or trees always exhibit strong elevation discontinuities. 
By iteratively minimizing the SSDGs, these discontinuities will be gradually flattened, hence resulting in inaccurately estimated $ z $-image at these elevation-transitioning regions.
Such discontinuities should be preserved during the super-resolution process.
Thus, a $ \mathcal{L}^1 $-norm term is additionally proposed in our approach to promote sparsity of the $ z $-image, i.e. to preserve the elevation discontinuities stemming from buildings and trees.

\paragraph{Propagation algorithm}
\begin{algorithm}[t]
	\caption{Solving Eq. \eqref{eq:zvtp1} by FISTA algorithm with constant step size $ \gamma $.}
	\label{alg:fista}
	\begin{algorithmic}
		\REQUIRE \mbox{ }
		
		\begin{itemize}[label=-]
			\itemsep0em
			\item sparse image $  \phi^{spa} $ ($  \phi^{spa}_{\Omega^*}=H_{\Omega^*}\psi^z $ or $  \phi^{spa}_{\Omega^*}=H_{\Omega^*}\psi^i $, $ \phi^{spa}_\Omega=0$)
			\item a maximum number of iterations $k_{max} $
			\item step size $ \gamma>0$
			\item soft thresholding parameter $ \lambda>0 $
			\item a tolerance value $\epsilon $ for stopping criterion
		\end{itemize}
		
		\COMMENT{$ k \leftarrow 1, t_0 \leftarrow 1, y^{(0)} \leftarrow \phi^{spa}$}
		
		\REPEAT 
		\STATE $ x^{(k)}_{\Omega}  = \mathcal{T}_{\lambda\gamma}{\left(y_{\Omega}^{(k-1)}-\gamma H_{\Omega}\nabla f_{\mathrm{SSDG}}\left(y^{(k-1)}\right)\right)}$ \\[3pt]
		\STATE $ t_k = \dfrac{1}{2}\left(1+\sqrt{1+4t_{k-1}^2}\right) $\\[3pt]
		\STATE $ y^{(k)}_{\Omega} = x^{(k)}_{\Omega} + \left(\dfrac{t_{k-1}-1}{t_k}\right)\times\left(x^{(k)}_{\Omega} -x^{(k-1)}_{\Omega} \right) $\\[3pt]
		\STATE $ k \leftarrow k+1 $\\[3pt]
		\UNTIL{$ k > k_{max} $ \OR $ \left|y^{(k)}-y^{(k-1)}\right| <\epsilon$}\\[3pt]
		
		\COMMENT{$ \widehat{\phi}_{\Omega^*}\leftarrow \phi^{spa}_{\Omega^*} $ \AND $  \widehat{\phi}_{\Omega}\leftarrow y^{(k)}_{\Omega}$}\\[3pt]
		\ENSURE dense image $ \widehat{\phi}$
	\end{algorithmic}
\end{algorithm}

The optimization problem described by Eq. \eqref{eq:zvtp1}, containing the term $ \left\|\phi\right\|_{1} $, is solved iteratively. 
Each iteration involves calculating the gradient descent of the SSDG term (i.e. $ \nabla f_{\mathrm{SSDG}} $) followed by a shrinkage/soft-threshold step. 
The shrinkage operator $  \mathcal{T}_{\alpha} : \mathbb{R}^n \rightarrow \mathbb{R}^n$ is defined as follows,
\begin{equation}\label{eq:shrinkage-operator}
		\mathcal{T}_{\alpha}(x) = (|x|-\alpha)_+ \times \mathrm{sign}(x)
\end{equation}
where  $ (|x|-\alpha)_+ = \max(|x|-\alpha, 0) $, and $ \alpha $ is a threshold value, which is set to $ \alpha = \lambda\gamma$ in Algorithm \ref{alg:fista}.

Algorithm \ref{alg:fista} presents the process of solving \eqref{eq:zvtp1}, using the Fast Iterative Shrinkage-Thresholding algorithm (FISTA) \cite{beck2009fast} with a constant step size.
In this Algorithm, the superscript $ {(k)} $ of a vector denotes its state at the $ k $-th iteration. 
The sub-vector $ x_{\Omega^*} $ (and $ y_{\Omega^*} $) contains only the values of pixels indexed by $ \Omega^* $, i.e. the pixels having a LiDAR transferred value. They remain unchanged during the propagation process. 
On the other hand, $ x_{\Omega} $ (and $ y_{\Omega} $) represents the sub-vector containing the values of pixels indexed by $ \Omega $, i.e. the null-valued pixels before the value propagation. 
The vector $ \phi $ without an index subscript is the vector contains all pixels, i.e. $ \phi = \phi_{\Omega \cup \Omega^*} $. 
For instance, $ \phi^{spa} $ represents the sparse image where pixels of index in $ \Omega^* $ are transferred from LiDAR data, while other pixels (i.e. the one of index from $ \Omega $) are null-valued.

\begin{figure*}[t]
	\centering
	\begin{subfigure}{0.46\linewidth}
		\includegraphics[trim=1cm 0cm 1cm 0.25cm,clip,width=\linewidth]{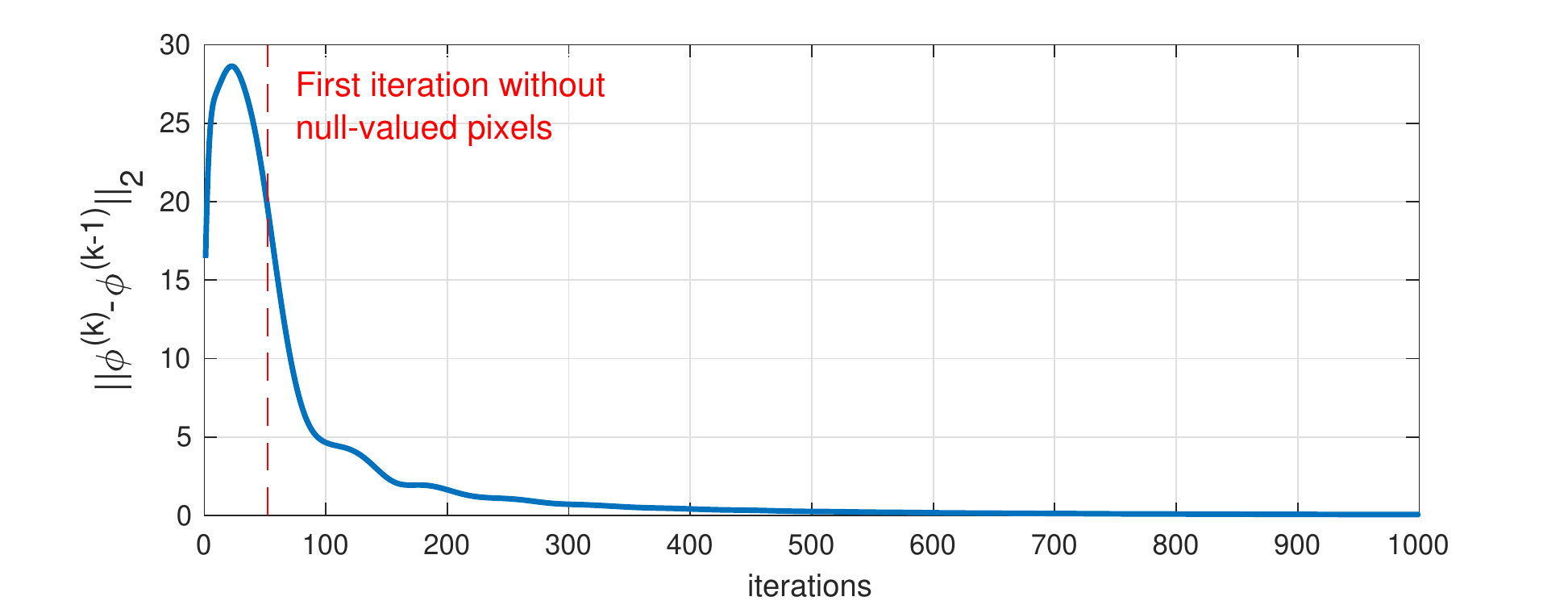}\caption{Error $\left\|\phi^{(k+1)}-\phi^{(k)}\right\|_{2} $}\label{subfig:error}
	\end{subfigure}
	\hspace{0.1cm}
	\begin{subfigure}{0.46\linewidth}
		\includegraphics[trim=1cm 0cm 1cm 0.25cm,clip,width=\linewidth]{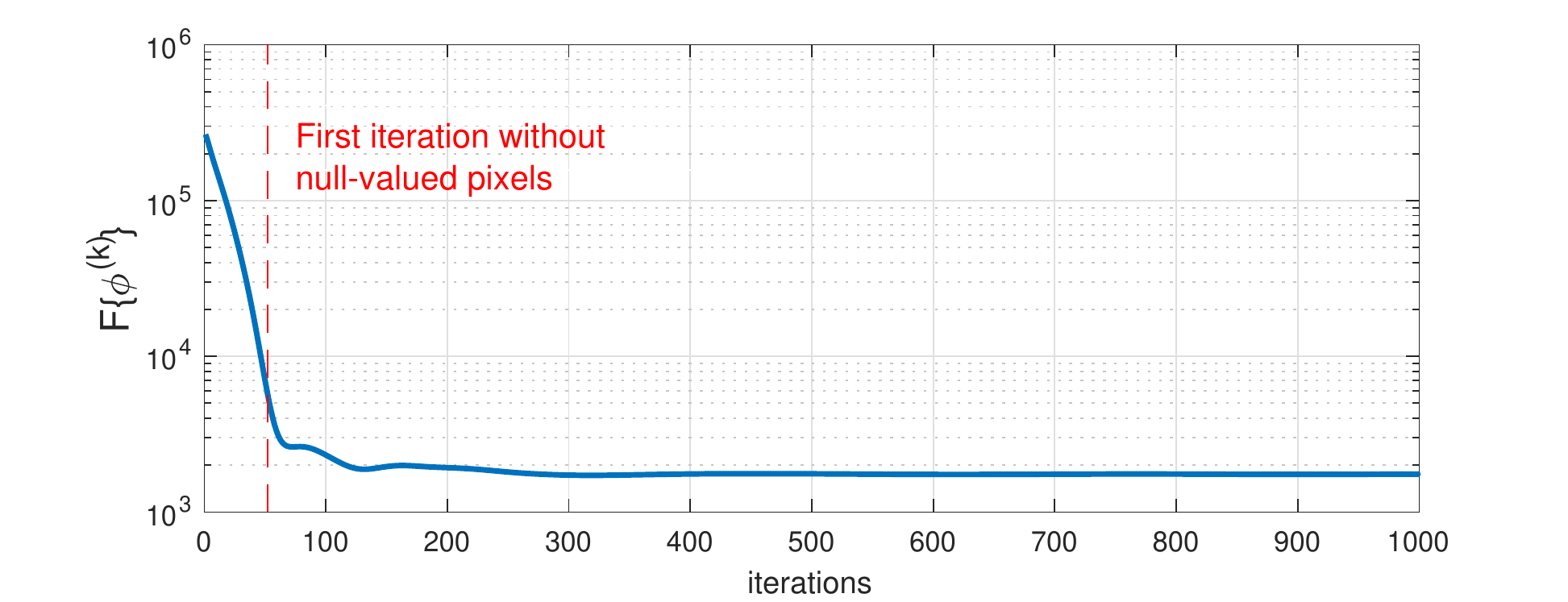}\caption{Function value $ \mathcal{F}(\phi^{(k)}) $ (in log-scale)}\label{subfig:function value}
	\end{subfigure}

	\vspace{0.15cm}
	\begin{subfigure}{0.46\linewidth}
		\includegraphics[trim=1cm 0cm 1cm 0.25cm,clip,width=\linewidth]{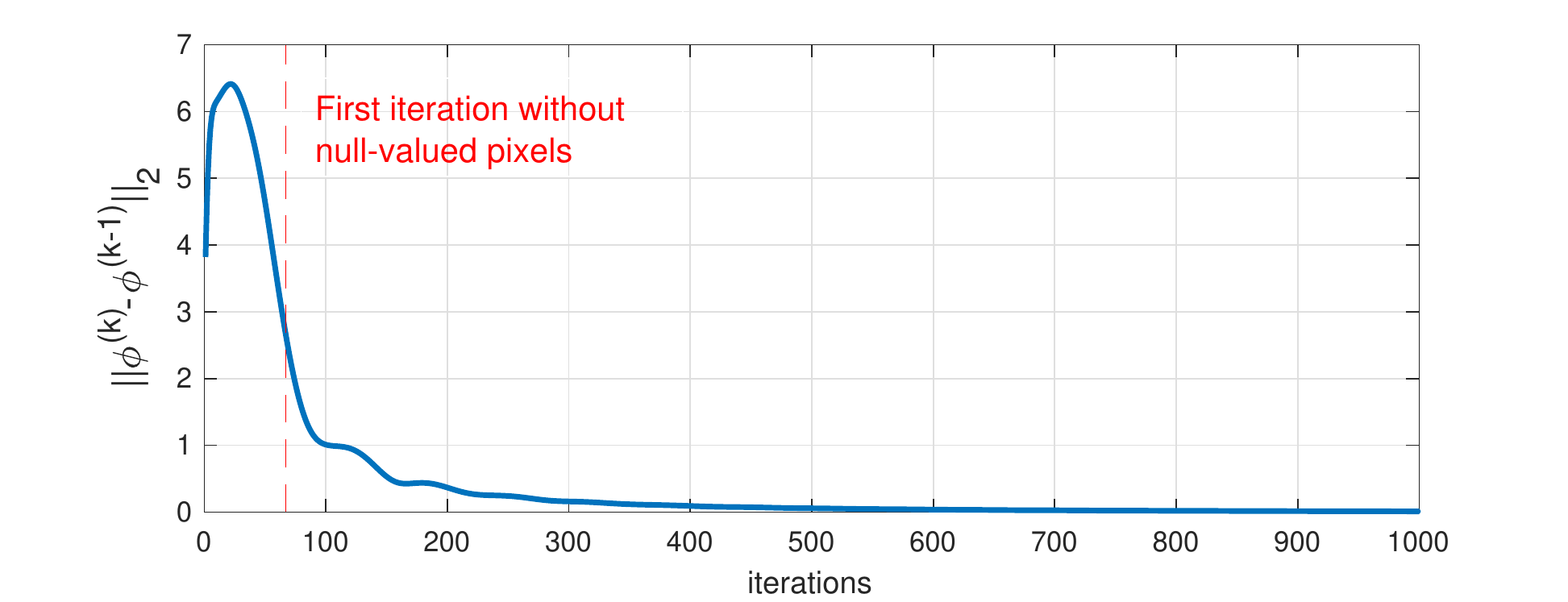}\caption{Error $\left\|\phi^{(k+1)}-\phi^{(k)}\right\|_{2} $}\label{subfig:error_int}
	\end{subfigure}
	\hspace{0.1cm}
	\begin{subfigure}{0.46\linewidth}
		\includegraphics[trim=1cm 0cm 1cm 0.25cm,clip,width=\linewidth]{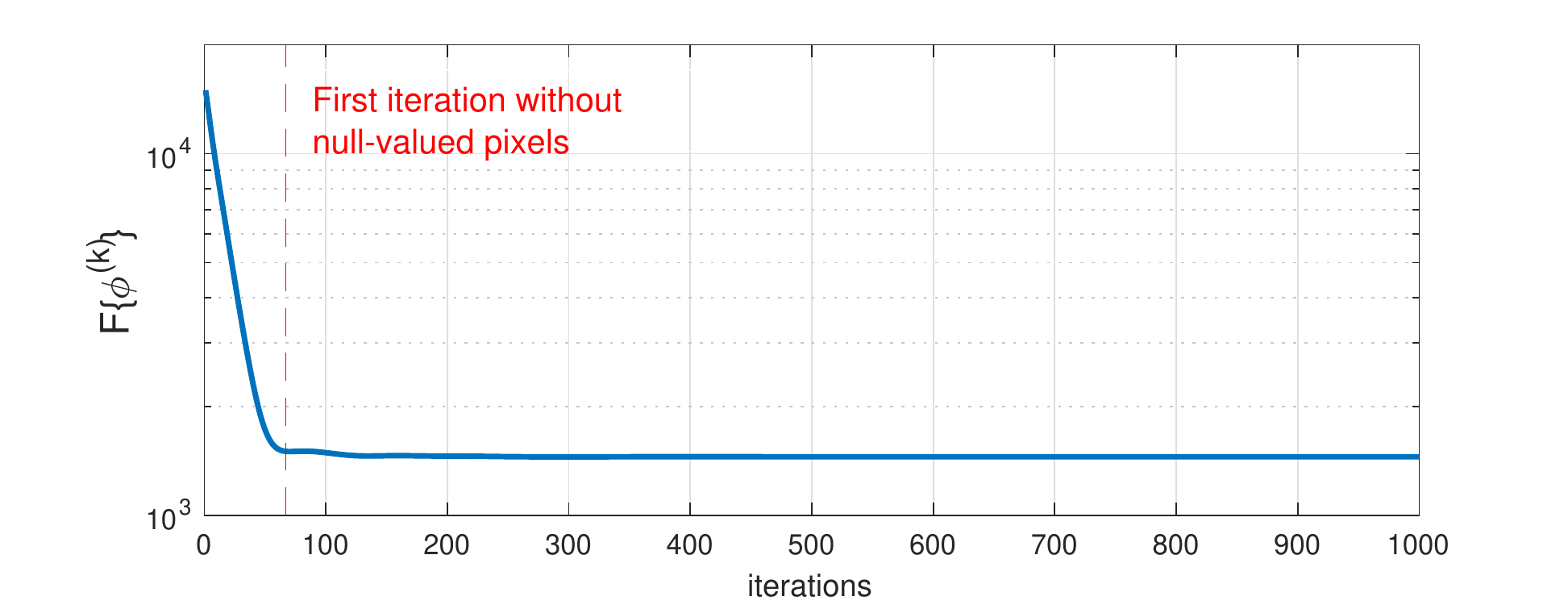}\caption{Function value $ \mathcal{F}(\phi^{(k)}) $ (in log-scale)}\label{subfig:function value_int}
	\end{subfigure}
	\caption{The errors $\left\|\phi^{(k+1)}-\phi^{(k)}\right\|_{2} $ and the cost values $ \mathcal{F}(\phi^{(k)}) $ plotted as a function of iterations. The vertical red-dashed lines indicate the first iteration where every pixel of the estimated image is filled. First row: the plots from the SR process of generating the $ z $-image; second row: the plots from the SR process of generating the $ i $-image.}
	\label{fig:errors}
\end{figure*}

\begin{figure*}[!t]
	\centering
	\begin{minipage}[b]{0.29\linewidth}
	\begin{subfigure}{\linewidth}
		\includegraphics[trim=0 0 0 0,clip,width=\linewidth]{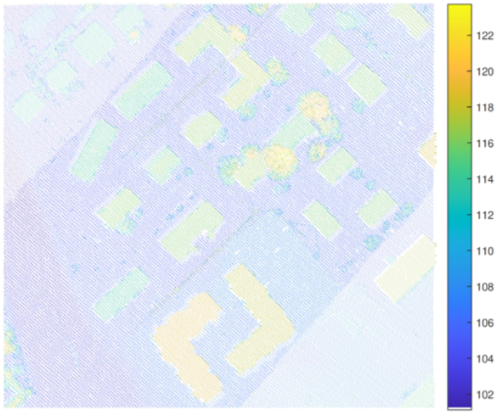}\caption{Sparse $ z $-image}\label{sfig:sparse-z-image}
	\end{subfigure}
	\begin{subfigure}{\linewidth}
		\includegraphics[trim=0 0 0 0,clip,width=\linewidth]{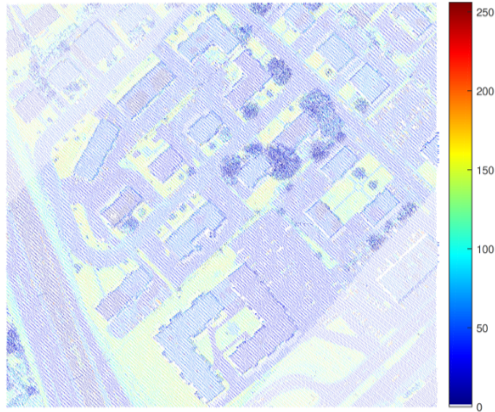}\caption{Sparse $ i $-image}\label{sfig:sparse-i-image}
	\end{subfigure}
	\end{minipage}
	\hspace{0.5cm}
	\begin{minipage}[b]{0.29\linewidth}
	\begin{subfigure}{\linewidth}	
		\includegraphics[trim=0 0 0 0,clip,width=\linewidth]{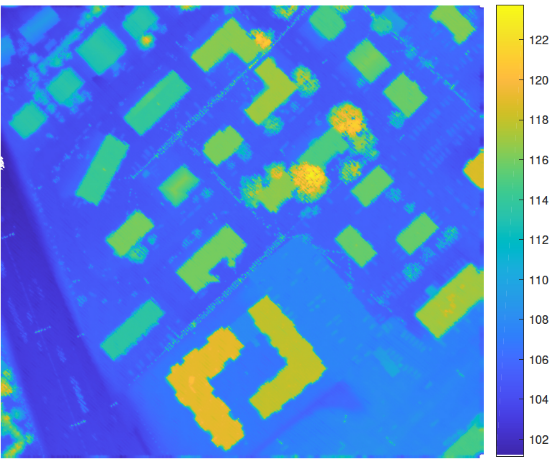}\caption{Dense $ z $-image}\label{sfig:z-image}
	\end{subfigure}
	\begin{subfigure}{\linewidth}
		\includegraphics[trim=0 0 0 0,clip,width=\linewidth]{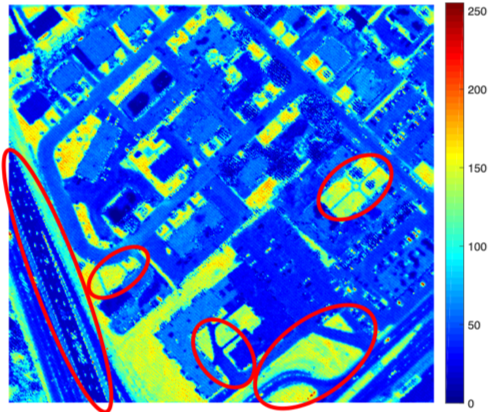}\caption{Dense $ i $-image}\label{sfig:i-image}
	\end{subfigure}
	\end{minipage}
	\hspace{0.5cm}
	\begin{minipage}[b]{0.29\linewidth}
		\begin{subfigure}{\linewidth}
		\includegraphics[trim=0cm 0cm 0cm 0cm,clip,width=0.85\linewidth]{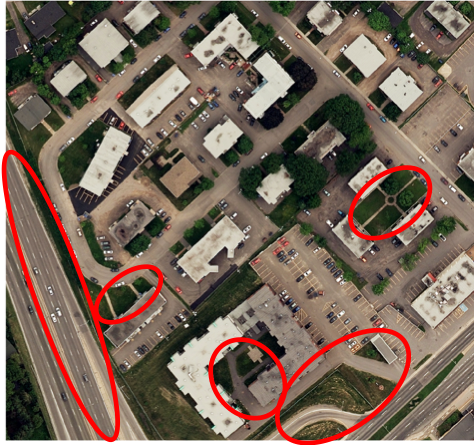}\caption{Optical image}\label{sfig:opt-image}
	\end{subfigure}
	\end{minipage}
	\caption{Illustration of the super-resolution results. First column: the sparse $ z $-image and $ i $-image from the value transfer process; second column: the dense $ z $-image and $ i $-image from the value transfer and propagation process; third column: the optical image of the same scene shown for the sake of comparison.}
	\label{fig:sparse_dense}
\end{figure*}

FISTA with its computational simplicity is adequate for solving large-scale problems.
It also converges more quickly than ISTA, with a rate of $ O(1/k^2) $ \cite{beck2009fast}. 
The convergence rate of the SRs is depicted in Fig. \ref{fig:errors}. 
Indeed, Fig. \ref{subfig:error} and \ref{subfig:error_int}, respectively, depict the errors between the estimated $ z $-images and $ i $-images at two consecutive iterations, i.e. $ \left\|\phi^{(k+1)}-\phi^{(k)}\right\|_2 $. 
The values of the cost function $ \mathcal{F}(\phi^{(k)}) $ through iterations are also shown in Fig. \ref{subfig:function value} and \ref{subfig:function value_int}. 
We can remark that after approximately 600 iterations, the estimated $ z $-image and $ i $-image have nearly converged into stable solutions.

Lastly, Fig. \ref{fig:sparse_dense} shows the results of a transfer and propagation of altitude and intensity values from the LiDAR data onto the frame of the optical image. The value transfer results are depicted through the sparse images (Fig. \ref{sfig:sparse-z-image} and \ref{sfig:sparse-i-image}), while the value propagation results are shown by the dense images (Fig. \ref{sfig:z-image} and \ref{sfig:i-image}). 
On the $ z $-images, the pixel color represents the altitude in meters.  
In contrast, the pixel color on the $ i $-images represents the intensity value between 0 and 255. 
The reference optical image on the same urban scene (Fig. \ref{sfig:opt-image}) allows a visual quality assessment of the super-resolved images. 
On the one hand, we can observe that the elevation of buildings and other off-terrain objects (e.g. trees, power lines), as well as the relief of the urban scene are well presented on the dense $ z $-image (Fig. \ref{sfig:z-image}), and correspond with the information in the optical image. 
On the other hand, different elements of the scene like buildings, grasses, or roads can be discriminated on the $ i $-image, similarly on the optical image. 
For example, several regions with distinctive elements are highlighted by the red ellipses on Fig. \ref{sfig:i-image} and  \ref{sfig:opt-image}.

\subsubsection{Estimation of local transformation model}\label{sssec:optim}
As aforementioned, an MI-based registration method involves many constraints. One of them is that the to-be-registered datasets must have the same resolution.
The solution to this problem is to use high-resolution LiDAR-based images generated by the presented SR approach.  

\paragraph{Proposed approach}
This paper presents an MI-based registration method that relies on the MI measured between the optical image and the $ i $-image. 
An NCMI-based registration method is also proposed. It involves measuring the NCMI between the optical image and both the  $ z $-image and $ i $-image.
Both similarity measures, MI and NCMI, are expected to achieve their maximum value when the images (i.e. the optical image and the LiDAR-based images) are geometrically aligned.

Moreover, the proposed fine registration also consists in a local approach of transformation model estimation.
It involves dividing the study area into many patches of same size and estimating a local transformation model for each patch. 
Such a patch-based approach allows accelerating the MI maximization process. 
For each patch of the optical image $ u_t $ and the LiDAR data $ \psi_t $ ($ 1\le t\le T $, with $ T $ is the total number of patches), the determination of the optimal set of camera pose parameters, denoted by $ \theta_t^* $, is carried out based on the maximization of MI or NCMI, as follows, 
\begin{equation}\label{eq:optim_mi}
\theta_t^*=\underset{\theta\in\Theta}{\arg\max}~\mathrm{MI}(f^i_{\mathrm{SR}}(\theta, \psi_t) ; u_t)
\end{equation}
\begin{equation}\label{eq:optim_ncmi}
\theta_t^*=\underset{\theta\in\Theta}{\arg\max}~\mathrm{NCMI}((f^i_{\mathrm{SR}}(\theta, \psi_t), f^z_{\mathrm{SR}}(\theta, \psi_t)) ; u_t)
\end{equation}
Eq. \eqref{eq:optim_mi} and \eqref{eq:optim_ncmi} present the maximizations based on, respectively, MI and NCMI. 
$ f^i_{\mathrm{SR}} $ and $ f^z_{\mathrm{SR}} $ represent the SR process that generates, respectively, the $ i $-image and $ z $-image (denoted by $ \phi $ from the previous sub-section \ref{sssec:sr}), given the camera pose parameters $ \theta $ and the LiDAR data $ \psi_t $.

Given two random variables $ A $ and $ B $ with  marginal probability distribution functions (pdf), $ p_A(a) $ and $ p_B(b) $ and joint pdf $ p_{AB}(a,b) $, the Mutual Information between $ A $ and $ B $, denoted by $ \mathrm{MI}(A;B) $, measures the degree of dependence of $ A $ and $ B $  by the distance between the joint pdf $ p_{AB}(a,b) $ and the pdf associated with the case of complete independence $ p_A(a).p_B(b) $. 
This entropic distance is expressed by the means of the Kullback-Leibler divergence measure \cite{vajda1989theory}, given by Eq. \eqref{eq:MI_theory},
\begin{equation}\label{eq:MI_theory}
\begin{array}{rl}
\mathrm{MI}(A;B) & = \sum\limits_{a,b} p_{AB}(a,b)\log\dfrac{p_{AB}(a,b)}{p_A(a).p_B(b)}\\[10pt]
& = H(A)+H(B)-H(A,B)
\end{array}
\end{equation}
where $ H(X)=-\sum_{x} p_X(x)\log p_X(x) $ is the \textit{Shannon entropy} of random variable $ X $. Its estimation is proposed by Mokkadem in \cite{mokkadem1989estimation}. 
The registration method based on the maximization of MI is originally introduced by \cite{maes1997multimodality}. 
Since then it has been extensively studied in many research areas, particularly to register an optical image with an image derived from LiDAR data. 
This image is either the LiDAR-derived DSM or the intensity image, which has the same resolution with the optical image \cite{parmehr2013automatic,Mastin2009}.

Another statistical similarity measurement used for the registration between LiDAR data and optical imagery (Eq. \eqref{eq:optim_ncmi}) is the Normalized Combined Mutual Information (NCMI) \cite{Parmehr2014}, given by Eq. \eqref{eq:ncmi}.  
\begin{equation}\label{eq:ncmi}
\mathrm{NCMI}((A,B);C)=\dfrac{H(A;B)+H(C)}{H(A;B;C)}
\end{equation}
NCMI-based registration method relies on the similarity between the optical image and both LiDAR images, i.e. DSM and intensity image which are inherently registered. This combined similarity measurement is shown to be more informative than the conventional MI \cite{cahill2010normalized}.

The authors of \cite{Mastin2009} compared the three usages of LiDAR-derived images in MI-based registration involving measuring its/their similarity with the optical image, i.e. \textit{(i)} using only the DSM image, \textit{(ii)} using only the intensity image, \textit{(iii)} using both images.
They demonstrate that the usage of the intensity image yields more accurate registration result than using the DSM image. 
The usage of both images is also shown to yield more accurate result than the two individual usages \cite{Parmehr2014}.

\paragraph{Implementation}
To resolve \eqref{eq:optim_mi} and \eqref{eq:optim_ncmi} we use Nelder-Mead simplex algorithm \cite{nelder1965simplex}.
Such algorithm is derivative-free and also straightforward in terms of implementation.
The initial values for the optimization are given by the $ \theta_{global} $, resulted from the coarse registration.
In this paper, the considered urban area is divided into equal patches, of which the size is chosen as 500 $ \times $ 550 pixels.
This patch size for the fine registration has been selected based on the study of \cite{parmehr2016automatic}.
It is not related to the building size. It only aims at reducing the computational cost while maintaining a sizable patch for a reliable MI calculation.
The division of the area into equal patches is irrespective of the distribution of buildings, or in other words, independent of the distribution of correspondences used in the coarse registration.

\subsubsection{Smoothing of patch-based registration results}\label{sssec:smoothing}
\begin{figure}[h]
	\centering
	\begin{subfigure}{0.4\linewidth}
		\centering\includegraphics[trim=2cm 2.25cm  2cm 2cm,clip,height=3cm]{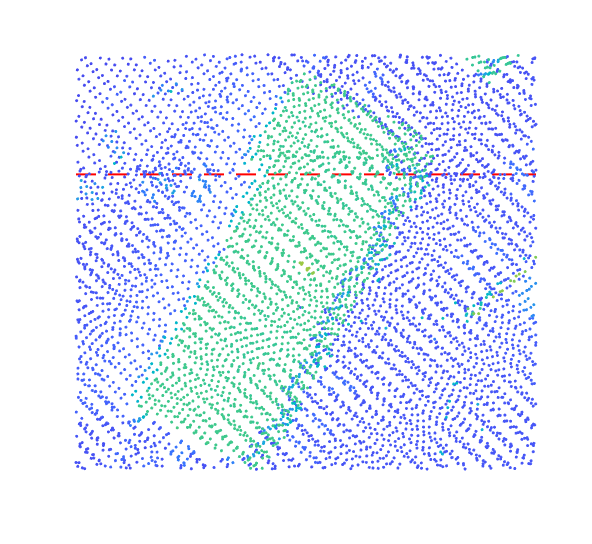}\caption{Before smoothing}\label{sfig:problem1_patch}
	\end{subfigure}
	\hspace{0.2cm}
	\begin{subfigure}{0.4\linewidth}
		\centering\includegraphics[trim=2cm 2cm  2cm 2cm,clip,height=3cm]{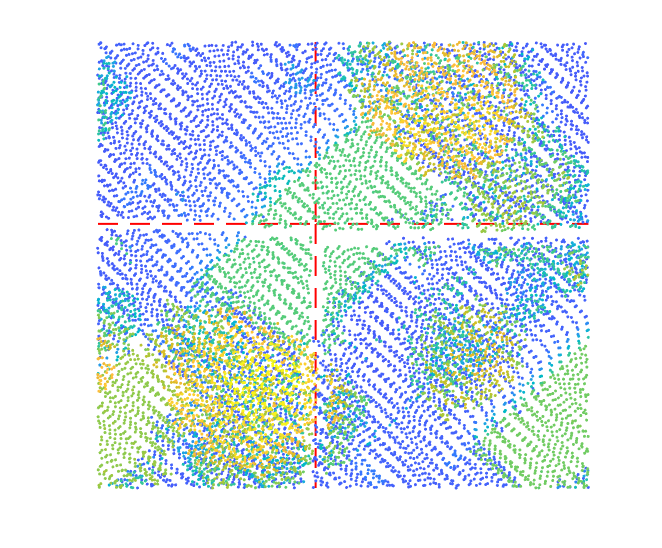}\caption{Before smoothing}\label{sfig:problem2_patch}
	\end{subfigure}

	\vspace{0.15cm}
	\begin{subfigure}{0.4\linewidth}
		\centering\includegraphics[trim=2cm 2.25cm  2cm 2cm,clip,height=3cm]{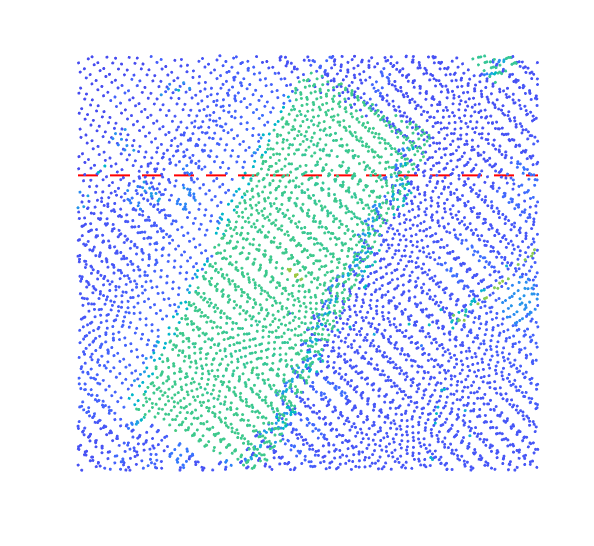}\caption{After smoothing}\label{sfig:problem1_patch_solved}
	\end{subfigure}
	\hspace{0.2cm}
	\begin{subfigure}{0.4\linewidth}
		\centering\includegraphics[trim=2cm 2cm  2cm 2cm,clip,height=3cm]{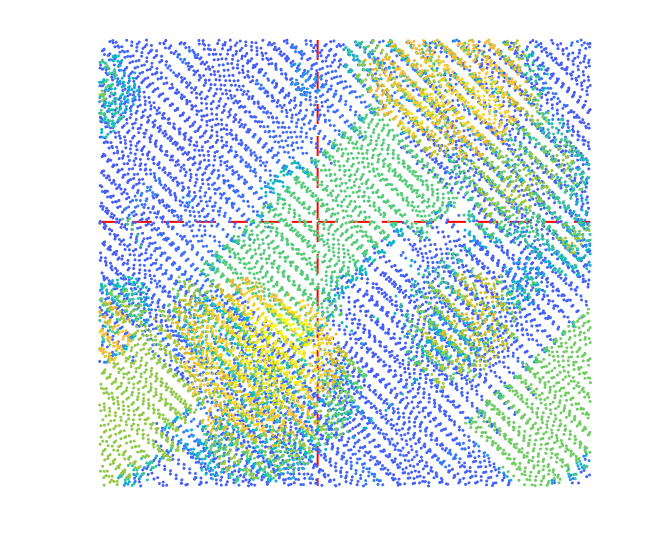}\caption{After smoothing}\label{sfig:problem2_patch_solved}
	\end{subfigure}

	\vspace{0.15cm}
	\begin{subfigure}{0.4\linewidth}
		\centering\includegraphics[trim=1.5cm 0.5cm  1.5cm 0.5cm,clip,height=3cm]{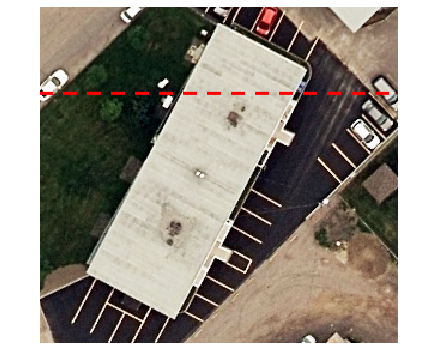}\caption{Optical image}\label{sfig:problem1_ref}
	\end{subfigure}
	\hspace{0.25cm}
	\begin{subfigure}{0.4\linewidth}
		\centering\includegraphics[trim=1.5cm 0.5cm  1.5cm 0.25cm,clip,height=3cm]{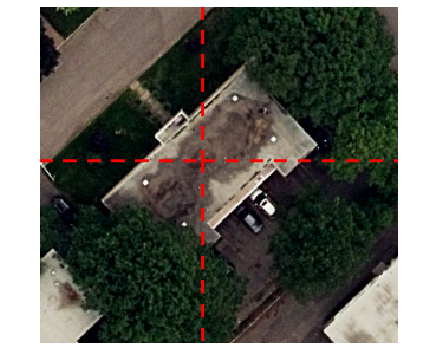}\caption{Optical image}\label{sfig:problem2_ref}
	\end{subfigure}
	\caption{
		Examples of \textit{(a)} a conflict, and \textit{(b)} a discontinuity of projected points between neighboring patches, and the results \textit{(c)} and \textit{(d)} on the same regions after the smoothing using IDW-based interpolation. 
		The red dashed lines represent the patch boundaries. 
		The optical images \textit{(e)} and \textit{(f)} of the considered areas are shown for the sake of comparison. 
		As such, the shapes of the exemplified buildings are well retrieved after the smoothing.}
	\label{fig:problems_patch}
\end{figure}

A potential problem expected from this patch-based registration approach is the incoherence of local camera pose parameters $ \theta^*_t $ between patches, especially in transition regions between patches.
For instance, Fig. \ref{fig:problems_patch} illustrates two examples where the difference of transformation models of patches causes a conflict (Fig. \ref{sfig:problem1_patch}) and a discontinuity (Fig. \ref{sfig:problem2_patch}) of projected points on the transition regions between patches. 
In these regions, there are 3-D points which are spatially neighboring but they belong to two adjacent patches. 
Consequently, they are projected with two different transformation models, hence causing such conflict or discontinuity.
However, such problem has not yet been addressed by existing work that presented similar patch-based approach \cite{parmehr2013automatic}.

In this paper, we present a smoothing of camera pose parameters from the patch-based transformation models. 
Instead of using the transformation model of a patch for the projection of every point in this patch, we propose to compute the projection of each 3-D point based on a weighted average of several neighboring local transformation models. 
We denote the center of neighboring patches by $ C_i$, and the local transformation model of the patches by $ \theta_i $, $ i=1,..,N $ where $ N=9 $ is the number of neighboring patches of a point $ p $.
The transformation model for the point $ p $, denoted by $ \theta(p) $, is interpolated using the Inverse Distance Weighting (IDW) \cite{shepard1968two} average of neighboring local transformation models, given by Eq. \eqref{eq:IDW}.
 \begin{equation}\label{eq:IDW}
 	\theta(p) = \left\lbrace
 	\begin{array}{ll}
	\dfrac{\sum_{i=1}^{N}{w_i \theta_i}}{{\sum_{i=1}^{N}}{w_i}}, & \textnormal{if } d(p,C_i)\ne 0\\[15pt]
	
	\theta_i,& \textnormal{if } d(p,C_i) = 0
 	\end{array}
 	\right.
 \end{equation}

 The weights $ w_i $ are computed by the inverse squared Euclidean distance from the considered point $ p $ to the neighboring patch centers $ C_i $, as follows,
\begin{equation}\label{eq:weights}
  		w_i=\frac{1}{d(p,C_i)^2}, i=1,..,N
\end{equation}
 Fig. \ref{sfig:problem1_patch_solved} and \ref{sfig:problem2_patch_solved} depict the outcomes of the resolved incoherence problem  between patches, using
 the IDW-based interpolation of patch-based camera pose parameters.

\section{Results and Discussion}\label{sec:results}
\subsection{Assessment methodology}
Experiments have been carried out to evaluate the quality of the registration and determine whether it is good enough to be beneficial for a subsequent data fusion or other applications. 
However, the lack of a ground truth, i.e. true values of the camera pose parameters, makes such an evaluation difficult. 
To overcome this problem, Mastin \textit{et al.} \cite{Mastin2009} proposed to use expert-chosen control points to determine these values. 
Otherwise, without a ground truth, the registration quality of existing methods has been assessed in these following manners:
\begin{enumerate}[label=\textit{(\roman*)}]
	\item Using a subjective quality indicator or by a visual assessment: e.g. a \textit{good assignment of 3-D point-to-pixel on the colorized point cloud} \cite[Sec. 3]{vo2016processing}, or assessing whether \textit{the images are close enough for the projective texture mapping }\cite{Mastin2009}, or based on how well the representations of objects (e.g. buildings, vegetation) align. 
	\item Using the average spatial discrepancy between datasets measured at manually determined check points, or using check pair lines. 
	\item Involving a determination of an optimal set of parameters that minimizes a cost function or maximizes a statistical dependency measurement. In other words, a registration is considered successful when the determined parameters are optimal.
	The cost function can be the MI or its variation between the optical image and the LiDAR intensity image \cite{Parmehr2014,parmehr2013automatic,Mastin2009}. It can also be defined by the pixel-wise distances calculated between the hyperspectral image and the LiDAR-derived image  \cite{brell2016improving}.
\end{enumerate}

Since a thorough quality assessment of a registration method is still missing, we present multiple evaluations in this paper. 
Firstly, a visual assessment is carried out based on the alignments of scene elements.
Secondly, an evaluation of building candidate extraction and matching steps from the coarse registration is carried out. 
Thirdly, since our proposed registration method already involves a maximization of MI between datasets for determining optimal camera pose parameters,
we perform subsequently two spatial discrepancy evaluations. They are based on \textit{(i)} check points which are the centroids of manually determined building roofs, and \textit{(ii)} check pair lines manually sketched from the two datasets.

\begin{figure}[t]
	\centering
	\begin{subfigure}[b]{0.47\linewidth}
		\centering\includegraphics[trim=0.25cm 0cm 0.25cm 0cm,clip,width=3.5cm]{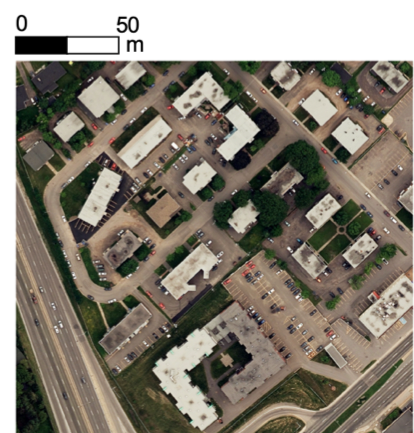}\caption{Urban area 1}\label{sfig:reg2}
	\end{subfigure}
	\hspace{0.01cm}
	\begin{subfigure}[b]{0.485\linewidth}
		\centering\includegraphics[trim=0cm 0cm 0cm 0cm,clip,width=4.25cm]{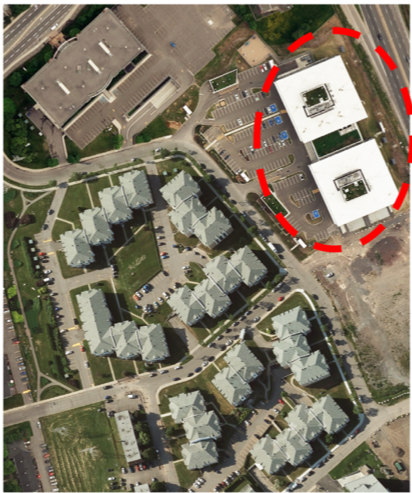}\caption{Urban area 2}\label{sfig:reg3}
	\end{subfigure}
	\caption{Two selected urban areas on which we evaluate the registration method.
	The first urban area is composed of 28 buildings; whereas the second urban area is composed of 12 buildings. 
	Particularly in the area 2, there are two buildings highlighted in red-dashed ellipse which were not built in 2011, namely the year when one of the LiDAR dataset was recorded.}
	\label{fig:regions}
\end{figure}

In this paper, we perform and assess the proposed registration method on four pairs of LiDAR-optical image datasets. 
They consist of four datasets on Quebec City (QC, Canada), namely two optical imagery and two LiDAR datasets.
It is worth reemphasizing that all four datasets were acquired from different platforms with different acquisition configurations at different times and have different spatial resolution.
Their specifications have been presented in Table \ref{tab:specs}. 
On the one hand, the first optical imagery dataset consists of the 15-cm resolution orthorectified aerial images acquired in 2016. 
The second dataset is provided by the Pléiades satellite involving 50-cm resolution images resulted from a pansharpening \cite{rahmani2010adaptive} of 50-cm resolution PAN images and 2-meter resolution MS images acquired in 2015.
On the other hand, the LiDAR datasets involve the airborne LiDAR point clouds acquired in 2011 and 2017. 
Their point spacing are, respectively, 70 cm and 35.4 cm.
We specifically measure the spatial discrepancy between LiDAR and imagery datasets, before and after carrying out our proposed registration method on two urban areas of representative characteristics in Quebec City.

The two selected urban areas are shown in Fig. \ref{fig:regions}.
The first area is composed of mostly planar roof buildings, which are relatively near each other.
In contrast, the second area is composed of big buildings and gable-roof houses. 
Particularly in the second area, there are two buildings located at the top left corner of Fig. \ref{sfig:reg3} (in the red-dashed ellipse) that were not built before 2013, hence they did not exist in the LiDAR dataset recorded in 2011. 
This absence highlights a temporal variability presented on this area.
As one can see, these two areas are composed of buildings of various types, sizes and shapes. 
Many buildings have inhomogeneous color rooftops, or even have small objects on them.
Several buildings are also surrounded by trees, or have casted very contrasting shadow regions (e.g. the gable houses on the second area).
These elements show the complexity of the selected areas. 
The versatility of the proposed method can be demonstrated through such different datasets on these complex areas.
Therefore, we are convinced that they are representative for the scenes which this research work aims to address.

\subsection{Visual assessment of the registration}
\begin{figure}[t]
	\centering
	\begin{subfigure}{\linewidth}
		\centering\includegraphics[trim=0cm 0.45cm 0cm 0.25cm,clip,width=8.3cm]{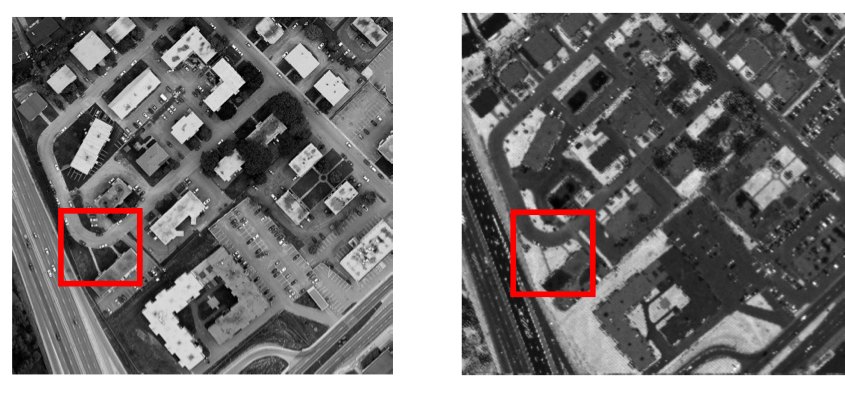}\caption{Optical image and $ i $-image on the area 1}\label{sfig:visual_assess1}
	\end{subfigure}

	\vspace{0.15cm}
	\begin{subfigure}{0.485\linewidth}
		\centering\includegraphics[trim=0cm 0cm 0cm 0cm,clip,width=3.9cm]{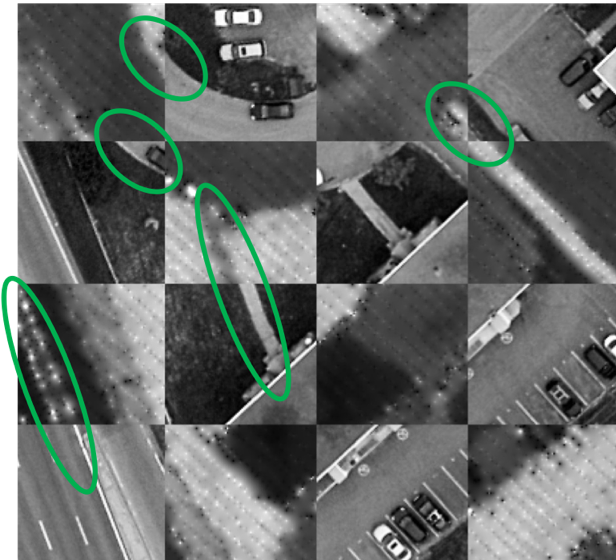}\caption{Checkerboard overlay (before registration)}\label{sfig:visual_assess2a}
	\end{subfigure}
	\hspace{0.005cm}
	\begin{subfigure}{0.485\linewidth}
		\centering\includegraphics[trim=0cm 0cm 0cm 0cm,clip,width=3.9cm]{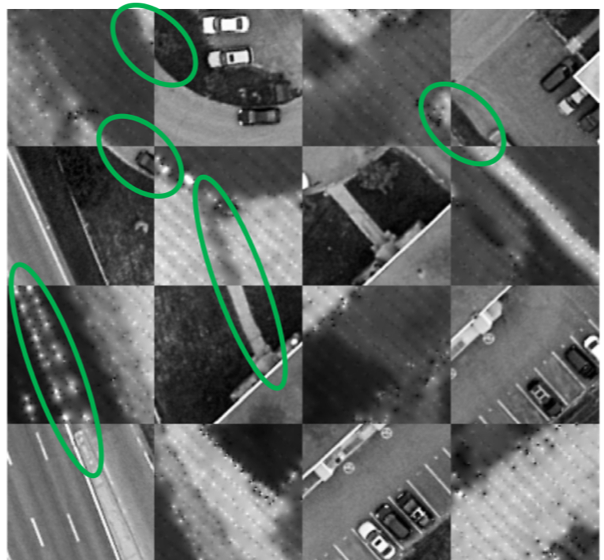}\caption{Checkerboard overlay (after registration)}\label{sfig:visual_assess2b}
	\end{subfigure}
	\caption{Visual assessment of the proposed registration method, focusing on the sub-image indicated by the red rectangle on the optical image and the $ i $-image from LiDAR data. The two sub-images are overlaid in a checkerboard-type display, in order to assess the alignment between multiple elements (green ellipses) yielded by the registration.}
	\label{fig:}
\end{figure}

Fig. \ref{sfig:visual_assess1} depicts a pair of images of the first urban area; the first one is the orthorectified optical image in 2016 (in gray-scale) and the second one is the $ i $-image from the LiDAR 2011 related to the proposed registration method.
Since the two images are of the same resolution, we display them in a checkerboard overlay, in order to assess visually the accuracy of the registration focusing on the small region encompassed by the red rectangles.
Fig. \ref{sfig:visual_assess2a} shows the  checkerboard overlay between the optical image and the $ i $-image before the registration, whereas Fig. \ref{sfig:visual_assess2b} shows the one after the registration.
It is worth noting a difference between these two images concerning the grayscale of different objects.
Indeed, on the optical image roads, pathways and buildings have bright pixels, whereas the grasses and trees have darker grayscale.
It is the opposite for the $ i $-image where roads, pathways and buildings are in darker grayscale than trees and grasses.

Let us focus on the alignment of multiple elements---continuous straight lines or curves from pathways and roads---on the region circled in green.
When putting together the two images that are well registered, these continuous elements should be visually aligned.
Indeed, being mindful of the grayscale differences between the two images, we can see that before the registration these circled elements are misaligned while all of them are rectified after the registration. 
It is also shown that the initial misalignments between the two datasets, even if relatively small, can still be reduced by the virtue of the proposed registration.

\subsection{Evaluation of building candidate extraction and matching steps from the coarse registration}

\begin{figure}[t]
	\centering
	\begin{subfigure}{0.45\linewidth}
		\centering\includegraphics[trim=0cm 0cm 0 0.25cm,clip,width=4.15cm]{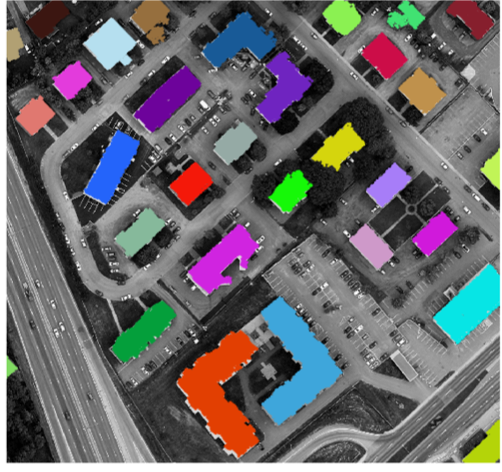}\caption{}
	\end{subfigure}
	\hspace{0.25cm}
	\begin{subfigure}{0.45\linewidth}
		\centering\includegraphics[trim=0cm 0cm 0cm 0.25cm,clip,width=4.15cm]{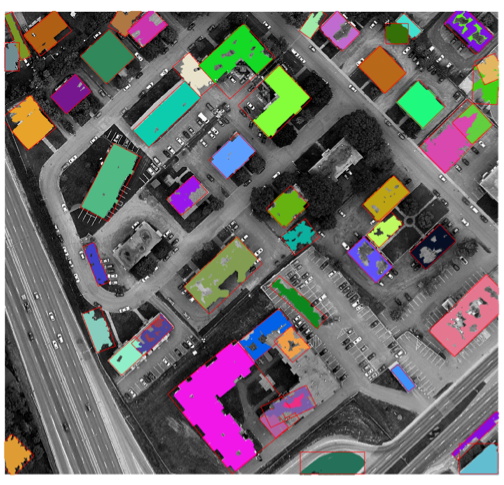}\caption{}\label{sfig:seg_to_compare1}
	\end{subfigure}
	
	\vspace{0.15cm}
	\begin{subfigure}{0.45\linewidth}
		\centering\includegraphics[trim=3.25cm 1cm 2.75cm 0.5cm,clip,width=4.15cm]{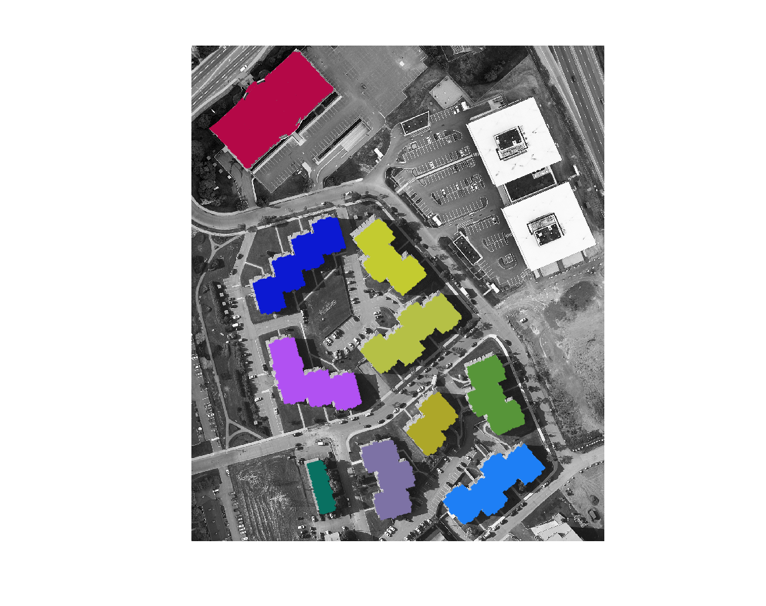}\caption{}
	\end{subfigure}
	\hspace{0.25cm}
	\begin{subfigure}{0.45\linewidth}
		\centering\includegraphics[trim=1.35cm 0.35cm 1.25cm 0.15cm,clip,width=4.15cm]{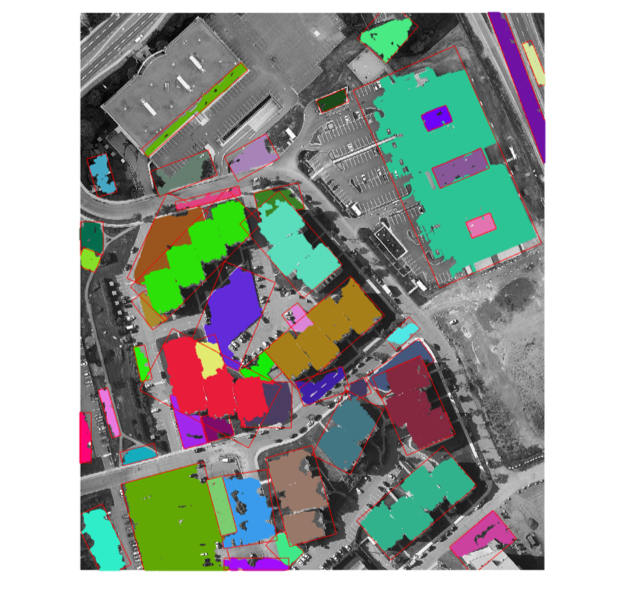}\caption{}\label{sfig:seg_to_compare2}
	\end{subfigure}

	\caption{Illustration of building extraction from LiDAR point cloud and from optical image. First column: LiDAR building segments, overlapped on optical image for visual comparison; Second column: image building segments with MBR filling percentage greater than $ 50\% $.}
	\label{fig:seg_to_compare}
\end{figure}

The results of the building extraction from LiDAR data (2011) and from aerial optical image (2016) are shown in Fig. \ref{fig:seg_to_compare}.
Table \ref{tab:comparative} summarizes the quantitative results of building extraction and matching on the selected areas. 
The  performance metrics are based on the number of true positives (TP), false alarms (FA), and misses (M). 
In the context of the building extraction, a TP indicates that a building is correctly extracted, whereas a FA represents a non-building segment incorrectly extracted as a building, and M means that a building exists but is not extracted. 
On the other hand, in the context of  segment matching, a TP indicates a good match, a FA represents an incorrectly matched pair of segments, whereas M means a pair of segments that should be matched are not paired. 
The Precision and Recall metrics are computed as follows, $ \textnormal{Precision}=\textnormal{TP}/{(\textnormal{TP}+\textnormal{FA})} $ and $ \textnormal{Recall}=\textnormal{TP}/{(\textnormal{TP}+\textnormal{M})} $.

\begin{table}[h]
	\centering
	\caption{Performance of building extractions and matching algorithms on selected areas}
	\begin{subtable}{\linewidth}
		\centering
		\renewcommand{\arraystretch}{1.25}
		\scalebox{\scaleone}{
			\begin{tabular}{c|P{0.16\linewidth}P{0.2\linewidth}|P{0.15\linewidth}P{0.15\linewidth}}
				\hline
				& Extracted from LiDAR data &  Extracted from image by mean shift & Matching result by RANSAC &  Matching result by GTM \\
				\hline
				{TP/FA/M} & 28/0/0 & 24/{21}/4 & 8/0/{12} & {19}/7/1 \\
				\hline
				{Precision} & 100\% & 53.33\% & 100\% & 73.08\% \\
				\hline
				{Recall} & {100\%} & {85.71\%} & 40\% & {95\%} \\
				\hline
			\end{tabular}
		}
		\vspace{0.05cm}
		\caption{On selected area 1 with 28 buildings in total.}
	\end{subtable}

	\vspace{0.15cm}
	\begin{subtable}{\linewidth}
	\centering	
	\renewcommand{\arraystretch}{1.25}
	\scalebox{\scaleone}{%
		\begin{tabular}{c|P{0.16\linewidth}P{0.2\linewidth}|P{0.15\linewidth}P{0.15\linewidth}}
				\hline
				& Extracted from LiDAR data &  Extracted from image by mean shift & Matching result by RANSAC &  Matching result by GTM \\
				\hline
				{TP/FA/M} & 10/0/0 & 11/37/1 & 7/1/2 & 9/1/0 \\
				\hline
				{Precision} & 100\% & 22.92\% & 87.5\% & 90\% \\
				\hline
				{Recall} & {100\%} & {91.67\%} & 77.78\% & {100\%} \\
				\hline
		\end{tabular}
	}
	\vspace{0.05cm}
	\caption{On selected area 2, there are in total 12 buildings (after 2013), but  there were only 10 (before 2013).}
	\end{subtable}
	\label{tab:comparative}
\end{table}

As shown by Table \ref{tab:comparative}, the building extraction from LiDAR data (both 2011 and 2017) has achieved 100\% of precision and recall thanks to the contribution of 3-D information. 
Similar performance of extracting buildings from LiDAR data has been found consistently, as presented by \cite{nguyen2019unsupervised2} when conducting tests on the ISPRS Vaihingen benchmark dataset \cite{cramer2010dgpf}.

On the other hand, the mean shift segmentation and subsequent segment refinement have yielded a high number of TPs but also a high number of FAs (i.e. twenty-one segments are incorrectly extracted as buildings). 
This number is even higher on the second area, i.e. thirty-seven FAs.
However, as one can see, the number of undetected building (i.e. M) is relatively low, namely four buildings on the first area and only one building on the second.
This results in the high Recall metric value from mean shift segmentation on both areas, 85.71\% and 91.67\%.

Considering the matching step, despite yielding relatively high precision (i.e. 100\% on the first area and 87.5\% on the second one), RANSAC provides a very low number of TPs and a high number of misses. GTM outperforms RANSAC on both areas, yielding more correct matches of building segments. 
However, on the first area, GTM yields a 95\% of recall, with a relatively high number of FAs (i.e. seven segment pairs are wrongly matched).  
These FAs are then eliminated by the subsequent validation based on segment area and direction (cf. Fig. \ref{fig:GTM}).

\subsection{Patch-based transformation model estimation}

\begin{figure*}[!t]
	\centering
	\begin{subfigure}{0.37\linewidth}
		\centering\includegraphics[trim=0.25cm 1cm  0.25cm 0.5cm,clip,width=\linewidth]{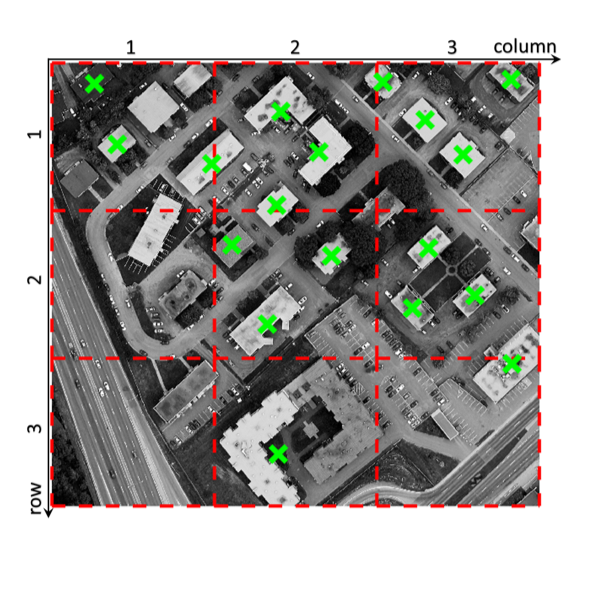}\caption{Patch division}\label{sfig:division_patch}
	\end{subfigure}
	\hspace{0.15cm}
	\begin{subfigure}{0.37\linewidth}
		\includegraphics[trim=0.25cm 1cm 0.25cm 0.5cm,clip,width=\linewidth]{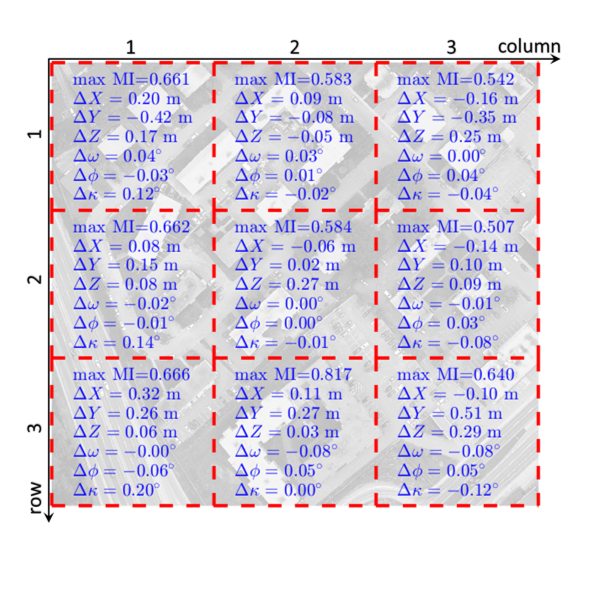}\caption{Parameter variations (i.e. $ \theta^{*}_t - \theta_{global} $)}\label{sfig:variation}
	\end{subfigure}
	\caption{Illustration of the local transformation model estimation. (a) Division of a considered urban area into equal patches. The red dashed lines depict the patch boundaries, whereas the green \textquoteleft x\textquoteright~represent the correspondences (i.e. matched primitives) from the coarse registration. (b) The values displayed in each patch are: first row: the maximized MI value; other rows: variation of the camera external parameters (i.e. $ \theta^{*}_t - \theta_{global} $).}
	\label{fig:local}
\end{figure*}

The division of a considered urban area into equal patches, and the local transformation model estimation are shown by Fig. \ref{fig:local}. 
On each patch, the maximized value of MI between the optical imagery and LiDAR data is displayed, as well as the variations of $ \theta^*_t $ compared to $ \theta_{global} $.
It should be noted that there is no relationship between the maximized MI values and the number of correspondences in each patch.
For example, the patch (2,1) with no correspondence can have a higher maximized MI value results than the patch (1,3) with four correspondences.
The relative difference among the maximized MI values stems from the different content of each patch.
These variations of $ \theta^*_t $ compared to $ \theta_{global} $ are different from one patch to another without any noticeable common pattern. 
Thus, potential incoherences between them can be expected. Such incoherences are resolved as a result of the IDW-based smoothing presented in \ref{sssec:smoothing}.

Table \ref{tab:local_global} summarizes the evolution of the resulting MI measurements between the global transformation model (i.e. outcome of the coarse registration) and the local transformation model (i.e. outcome of the fine registration) on each individual patch. 
Taking into account the number of correspondences among patches (as shown in Fig. \ref{sfig:division_patch}), we remark lower MI gains (not maximized MI values) for the patches with three to four correspondences than for the patches with fewer correspondences. 
In fact, the higher MI gains on the patches with few or no correspondence (the bold rows in Table \ref{tab:local_global}) shows the interest of the fine registration. Indeed, the coarse registration on these patches is less effective than on the patches with many correspondences, requiring the fine registration step to compensate for more data misalignment, hence resulting in higher MI gains.

\begin{table}[h]
	\renewcommand{\arraystretch}{1.2}
	\centering
	\caption{Evolution of MI from the coarse registration to the fine registration. The boldface rows highlight the patches having zero or one correspondence.}
	\scalebox{0.9}{
		\begin{tabular}{ccccc}
			\hline
			Patch & \multirow{2}{1.5cm}{\# of correspondences} & \multirow{2}{*}{$ \mathrm{MI}(\theta_{global}) $} & \multirow{2}{*}{$ \mathrm{MI}(\theta^*_{t}) $} & \multirow{2}{*}{$\dfrac{\mathrm{MI}(\theta^*_{t})}{\mathrm{MI}(\theta_{global})} -1$} \\
			(row, column) & & & & \\
			\hline
			(1,1)  & 3 & 0.615 & 0.661 & 7.4\% \\
			(1,2) & 3 & 0.546 & 0.583 & 6.7\% \\
			(1,3) & 4 & 0.479 & 0.542 & 13.1\% \\
			\textbf{(2,1)} & \textbf{0} & \textbf{0.572} & \textbf{0.662} & \textbf{15.8\%} \\
			(2,2) & 3 & 0.532 & 0.584 & 9.6\% \\
			(2,3) & 3 & 0.409 & 0.507 & 23.9\% \\
			\textbf{(3,1)}  & \textbf{0} & \textbf{0.393} & \textbf{0.666} & \textbf{69.4\%}\\
			\textbf{(3,2)} & \textbf{1} & \textbf{0.700} & \textbf{0.817} & \textbf{16.7\%} \\
			\textbf{(3,3)} & \textbf{1} & \textbf{0.329} & \textbf{0.640} & \textbf{94.5\%} \\
			\hline
			Average & - & 0.504 & 0.629 & 24.8 \% \\
			\hline
	\end{tabular}}
	\label{tab:local_global}
\end{table}

\subsection{Spatial discrepancy evaluation}
The third evaluation focuses on assessing the registration based on the spatial discrepancy between the datasets. 
In this regard, we propose to use manually determined check points (i.e. centroids of building roofs) and check pair lines (i.e. mainly building straight boundaries). 
They are determined from the optical image, and from the high-resolution $ z $-image generated from LiDAR point cloud. 
Indeed, in order to estimate the spatial discrepancy of datasets before the registration, a $ z $-image is generated using the presented SR process with a vertical projection of LiDAR points. 
For the coarse registration, another $ z $-image is generated using the global transformation model given by $ \theta_{global} $. 
Finally, in order to assess the spatial discrepancy after the fine registration (either MI or NCMI), the smoothed patch-based transformation model is used to generate the $ z $-image.
In the two following assessments, we propose to use $ z $-image instead of $ i $-image for these assessments, since it allows a better manual determination of building roofs and building boundary line segments (cf. Fig. \ref{sfig:z-image} and \ref{sfig:i-image}).

\subsubsection{Based on centroids of building roofs}
The first evaluation is carried out based on check points which are the centroids of manually delineated building roofs on the two selected areas. 
The distance (in meters) between the centroids of each pair are measured.
A smaller distance indicates a more accurate registration.
Each column of Table \ref{tab:eval_centroids} presents the spatial discrepancy evaluation between one LiDAR dataset and one optical imagery dataset, among the four datasets described in Table \ref{tab:specs}.
The evaluation is presented by the mean and standard deviation of the measured distances.
Indeed, the assessments on the registration between the LiDAR data 2011 and then the LiDAR data 2017 with the orthorectified aerial imagery 2016 are given by the column one and two of Table \ref{tab:eval_centroids}. 
Then, the ones between these LiDAR datasets with the Pléiades multispectral imagery data 2015 are provided by the column three and four.  
The spatial discrepancy values between the two considered datasets are averaged on all check points from the two selected areas.
An insignificant difference of approximately 10-15 cm is obtained between the two areas.
The gain values are computed based on average spatial discrepancy values after registration with respect to the values before registration.

\begin{table*}[h]
	\centering
	\caption{Building region centroids-based spatial discrepancy evaluation.}
	\renewcommand{\arraystretch}{1.25}
	\scalebox{\scaleone}{%
		\begin{tabular}{c|ccc|ccc|ccc|ccc}
			\hline
			& \multicolumn{3}{c|}{\multirow{2}{0.18\linewidth}{\centering LiDAR data (2011) and aerial imagery (2016)}} & \multicolumn{3}{c|}{\multirow{2}{0.18\linewidth}{\centering LiDAR data (2017) and aerial imagery (2016)}} & \multicolumn{3}{c|}{\multirow{2}{0.2\linewidth}{\centering {LiDAR data (2011) and Pléiades imagery (2015)}}} & \multicolumn{3}{c}{\multirow{2}{0.2\linewidth}{\centering LiDAR data (2017) and Pléiades imagery (2015)}}\\
			& \multicolumn{3}{c|}{} &\multicolumn{3}{c|}{} &\multicolumn{3}{c|}{} & \multicolumn{3}{c}{} \\
			\hline
			&Mean&(Std)& Gain & Mean & (Std) & Gain & Mean & (Std) & Gain & Mean & (Std) & Gain \\
			\hline
			Before registration & 1.08&(0.52) & - & 1.05 &(0.68)  & - & 42.89 & (1.47) & - & 44.43 & (1.73) & -  \\
			\hline
			Coarse registration & 0.56&(0.30) &  48.15\% & 0.54 &(0.55) & 48.57\% & 2.06 & (1.24) & 95.20\% & 1.39 & (0.44) & 96.87\% \\
			\hline
			MI-based fine registration & 0.46&(0.29) & 57.41\% & 0.43 &(0.32) & 59.05\% & 1.41 & (0.78) & 96.71\% & 1.20  & (0.58) &97.30\% \\
			\hline
			NCMI-based fine registration & 0.40&(0.27) &  62.96\% & 0.35 &(0.31) & 66.67\% & 0.99 & (0.45) &97.69\% & 0.82 & (0.45) &98.15\% \\
			\hline
			Gain of using NCMI over MI & \multicolumn{3}{c|}{13.04\%} & \multicolumn{3}{c|}{18.60\%}& \multicolumn{3}{c|}{29.79\%}& \multicolumn{3}{c}{31.66\%}\\
			\hline
	\end{tabular}
	}
	\label{tab:eval_centroids}
\end{table*}

\paragraph{Between orthorectified aerial image and LiDAR data}
Considering the orthorectified aerial image (2016), as a result of the image orthorectification, the average discrepancy between this dataset and the airborne LiDAR dataset (both 2011 and 2017) is already relatively  small, i.e. respectively 1.08 and 1.05 meters.
The results summarized by Table \ref{tab:eval_centroids} show that our proposed registration yields an even smaller discrepancy.
Indeed, the proposed coarse registration method results in a reduction of these values by 48.15\%. 
This reduction highlights the effectiveness of repositioning the datasets closer to each other.
Then, a spatial discrepancy of 40 cm between the LiDAR data (2011) and the orthorectified aerial imagery (2016), and of 35 cm between the LiDAR data (2017) and the orthorectified aerial imagery (2016) are provided by the NCMI-based fine registration method.

It is worth noting that both the LiDAR datasets acquired in 2011 and 2017 involve a horizontal accuracy of approximately 17 cm; whereas the horizontal accuracy of the orthorectified aerial imagery is 16.5 cm (cf. Table \ref{tab:specs}).
It means that the resulting discrepancy values, respectively 40 and 35 cm, are only slightly bigger than the combination of horizontal accuracy of the considered datasets.

Also, the reported average discrepancy between the LiDAR data 2011 and the orthorectified aerial imagery 2016 after the registration (i.e. 40 cm) is slightly bigger than 1/2 of the average point spacing of the considered LiDAR point cloud (i.e. 70 cm). 
On the other hand, regarding the registration between between LiDAR data 2017 and the orthorectified aerial imagery 2016, the resulting average discrepancy (i.e. 35 cm) approximates the average point spacing of the LiDAR point cloud (i.e. 35.4 cm).

\paragraph{Between Pléiades image and LiDAR data}
The discrepancy evaluation of the registration between the airborne LiDAR data (2011 and 2017) and the Pléiades imagery (2015) can be analyzed similarly from the results presented at the third and fourth columns of Table \ref{tab:eval_centroids}. 
The resolution of the Pléiades optical image is 50 cm, and its horizontal accuracy is theoretically between 1 and 3 meters, depending on the usage of ground control points on the considered area. 
These two characteristics of the Pléiades imagery data, especially the horizontal accuracy, are the major factors causing its registration with the LiDAR data (both 2011 and 2017) to be not as accurate as the registration between the LiDAR data and the aerial image (2016).
When regarding the resulting average discrepancy (i.e. 0.99 and 0.82 meters) and taking into account the spatial resolution of the datasets (i.e. 50 cm for the Pléiades imagery data and 70 cm or 35.4 cm for the LiDAR point spacing), one may interpret that these results are not good enough. 
However, as the horizontal accuracy of the Pléiades imagery dataset varies between 1 to 3 meters as aforementioned, we would argue that such average registration discrepancy results are highly desirable.

Comparing the MI-based and the NCMI-based fine registration, a gain averaging approximately 23\% is achieved when using NCMI instead of MI. 
In other words, as it has already been highlighted by other studies \cite{Mastin2009,Parmehr2014}, we also demonstrated that using both $ z $- and $ i $-images yields more accurate result than using either one of them. 
However, it is worth noting the research context of the present study is more complex than those of the previous works on this topic.
Nevertheless, it is worth noting that the NCMI-based fine registration takes twice as long as the MI-based, since it involves performing the SR twice, for the $ z $-image and the $ i $-image. 
Therefore, the use of both statistical similarity measurements shows beneficial result but not seamlessly. 
And the decision that which one of them should be used is essentially a compromise between accuracy and computational cost of the registration method.

Overall, the proposed coarse-to-fine method yields a very significant reduction of the spatial discrepancy between datasets, namely 63\% to 67\% for the registration of LiDAR data and orthorectified aerial imagery, and 98\% for the registration of LiDAR data and Pléiades imagery.

\subsubsection{Based on check pair line segments}
\begin{figure}[t]
	\centering
	\begin{minipage}{\linewidth}
		\centering\includegraphics[trim=0cm 0cm  0cm 0cm,clip,width=5cm]{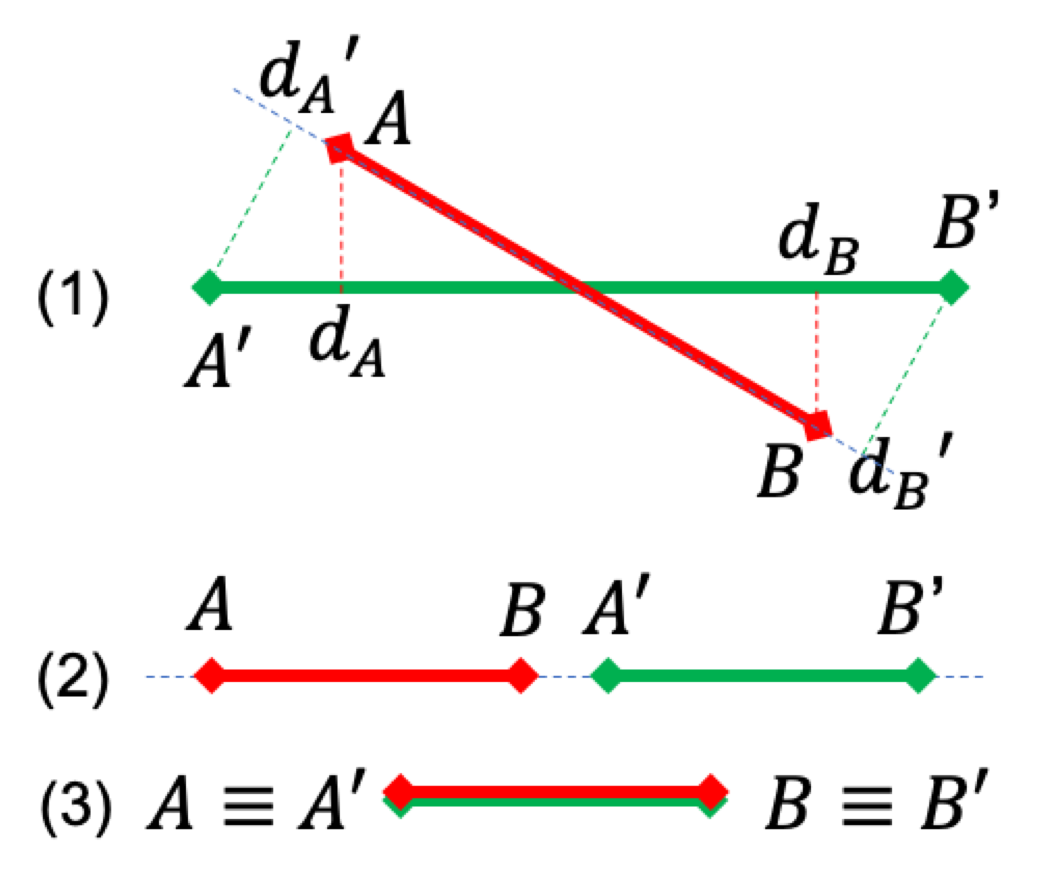}
		\label{sfig:CPL_illus}
	\end{minipage}
	
	\vspace{0.15cm}
	\begin{minipage}{\linewidth}
		\centering
		\scalebox{0.83}{
		\begin{tabular}{ccc}
			\hline
			& Distance measure \cite{peng2019automatic} & Modified distance measure \\\hline
			(1) Intersection & $ (d_A+d_B)/2 $ & $ \max\{d_A,d_B,d_A ',d_B '\}$ \\\hline
			(2) Collinear & 0 & $ \max\{|AA'|, |BB'|\} $ \\\hline 
			(3) Identical & 0 & 0 \\
			\hline
		\end{tabular}
		}
		\label{sfig:}
	\end{minipage}
	\caption{Illustration of the modified distance measure for the check pair line evaluation. Note the difference of the resulting distances between the collinear and the identical cases.}
	\label{fig:CPL_illus}
\end{figure}

The authors of \cite{peng2019automatic} proposed an evaluation based on distances between  line segments that are manually sketched from the two datasets. 
Compared with check point-based evaluation, this evaluation relies on a higher geometrical basis (i.e. line segments compared with points) to assess  the registration.
The distance proposed in \cite{peng2019automatic} between two line segments $ p $ (with $ A $ and $ B $ are its end-points) and $ q $ is defined as follows,
\begin{equation}\label{eq:cpl}
		d(p,q)=\dfrac{1}{2}(d_A+d_B)
\end{equation}
where $ d_A $ and $ d_B $ respectively denote the distances from $ A $ and $ B $ to the line segment $ q $. 
However, this distance given by Eq. \eqref{eq:cpl} is not quite relevant, because it yields small values when the two line segments are far away but collinear. 
Based on the literature review of line segment distances \cite{wirtz2016evaluation}, the Hausdorff line segment distance \cite{alt1995approximate} is more suitable for this evaluation.
It measures the longest of all the distances from a point on one line segment to the other segment, and it equals zero only if the two line segments are identical, i.e. same two end-points. 
Thus, it reflects fairly the discrepancy between two line segments, even when they intersect or are collinear. 
Fig. \ref{fig:CPL_illus} shows such improvement when using the Hausdorff distances instead of the Euclidean distances in three typical cases, namely intersection, collinear and lastly identical line segments. 

\begin{table*}[h]
	\centering
	\caption{Check pair line-based spatial discrepancy evaluation.}
	\renewcommand{\arraystretch}{1.25}
	\scalebox{\scaleone}{
		\begin{tabular}{c|ccc|ccc|ccc|ccc}
			\hline
			& \multicolumn{3}{c|}{\multirow{2}{0.18\linewidth}{\centering LiDAR data (2011) and aerial imagery (2016)}} & \multicolumn{3}{c|}{\multirow{2}{0.18\linewidth}{\centering LiDAR data (2017) and aerial imagery (2016)}} & \multicolumn{3}{c|}{\multirow{2}{0.2\linewidth}{\centering {LiDAR data (2011) and Pléiades imagery (2015)}}} & \multicolumn{3}{c}{\multirow{2}{0.2\linewidth}{\centering LiDAR data (2017) and Pléiades imagery (2015)}}\\
			& \multicolumn{3}{c|}{} &\multicolumn{3}{c|}{} &\multicolumn{3}{c|}{} & \multicolumn{3}{c}{} \\
			\hline
			&Mean&(Std)& Gain & Mean & (Std) & Gain & Mean & (Std) & Gain & Mean & (Std) & Gain \\
			\hline
			Before registration & 1.19 & (0.67) & - & 1.08 & (0.80) & - & 44.61 & (0.74) & - & {44.68} & (2.02) & -\\
			\hline
			Coarse registration & {0.95} & (0.83) & 20.17\%  & 0.81 & (0.32) & 25.00\% & 2.01 & (0.39) & 95.49\% & 2.18 & (1.00) & 95.12\% \\
			\hline
			MI-based fine registration& 0.64 & (0.30) & 45.82\% & 0.67 & (0.29) & 37.96\% & 1.95 & (0.65) &  95.63\% & 1.92 & (1.25)& 95.70\%  \\
			\hline
			NCMI-based fine registration & 0.64 & (0.29) & 45.95\% & 0.63 & (0.26) & 41.67\% & 1.47& (0.61) & 96.70\% & {1.21} & (0.33) & 97.29\%  \\
			\hline
			Gain of using NCMI over MI & \multicolumn{3}{c|}{2.33\%} & \multicolumn{3}{c|}{5.97\%}& \multicolumn{3}{c|}{24.62\%}& \multicolumn{3}{c}{36.98\%}\\
			\hline
	\end{tabular}
	}
	\label{tab:eval_CPL}
\end{table*}

In this assessment, two sets of seventy-two line segments are manually sketched on the optical image and on the generated $ z $-image of the two selected areas. They are then manually matched, yielding check pair line segments. 
The source for these line segments are mainly the building straight boundaries.
Then, the Hausdorff distance between each pair is computed. 
A smaller distance indicates a more accurate registration. 
Table \ref{tab:eval_CPL}  summarizes the quantitative results of check pair line-based evaluation of the registrations between airborne LiDAR data (2011 and 2017) with  orthorectified aerial imagery (2016), and with the Pléiades multispectral imagery data (2015).

As we can see from Table \ref{tab:eval_CPL}, the discrepancy between the datasets measured based on manually sketched line segments is significantly reduced after each of the registration, i.e. the coarse and the fine registration. 
On the one hand, the proposed registration method ultimately yields an average discrepancy of 0.63 and 0.64 meters between the LiDAR data (2017 and 2011) and the orthorectified aerial imagery (2016).  
On the other hand, for the registration of LiDAR data (2017 and 2011) and the Pléiades multispectral imagery (2015), the resulting check pair line-based discrepancy value is 1.21 and 1.47 meters. 
However, it is important to remind the principle of Hausdorff distance in order to evaluate the discrepancy results. 
For example, we consider two nearly identical horizontal line segments (i.e. parallel to $ x $-axis), having same first end-point, and the second end-point of one line segment is four pixels away from the second end-point of the other line segment (with a pixel size of 15 cm). 
Consequently, the resulting Hausdorff distance between them is 60 cm. 
Therefore, it should be noted that the mentioned discrepancy values yielded by our proposed method are relatively small. 
Overall, for all four registrations, the check pair line-based discrepancy varies between three and four pixels.

A discrepancy reduction of approximately 42\% to 46\% is achieved on the registration between the LiDAR data (2011 and 2017) and the orthorectified aerial imagery (2016). 
Similarly, a spatial discrepancy reduction of approximately 97\% (96.70\% and 97.29\%) is benefited from the registration between the LiDAR data (2011 and 2017) and the Pléiades multispectral imagery (2015). 

Finally, it can be noted that, in the registration between the LiDAR data and the Pléiades image, the benefit of using NCMI instead of MI is much more evident than in the registration between the LiDAR data and the orthorectified aerial image.
Indeed, based on both check point-based and check pair line-based evaluation result, using NCMI instead of MI results in average gains of 30.7\% and 30.8\% of spatial discrepancy reduction for the registration between the LiDAR data and the Pléiades image.  
The first percentage is computed from check point-based evaluation result  (i.e. 29.79\% and 31.66\% from Table \ref{tab:eval_centroids}), whereas the second percentage is computed from check pair line-based evaluation result (i.e. 24.62\% and 36.98\% from Table \ref{tab:eval_CPL}).
On the other hand, for the registration between the airborne LiDAR data  and the orthorectification aerial image, these average gains are only 15.8\% and  4.15\%.

Both spatial discrepancy assessments and all these mentioned elements show that the results yielded by our proposed method are relevant.
These presented assessments have also shown and validated the versatility of the proposed method, through the differences between the registered datasets and the complexity of the test areas.
However, it should be noted that it is virtually impossible for a registration method to perform well on any other scene without an adaptation or re-parametrization.
Notwithstanding, in another context, e.g. European urban scenes, the same registration approach should be applicable without major difficulties.

\section{Conclusions and Perspectives}\label{sec:conclusions}
This paper has presented and evaluated a coarse-to-fine registration method between airborne LiDAR data and optical imagery. It is dedicated to overcome the challenges associated with the difficult context, where the two datasets are not acquired from the same platform, neither from the same point of view nor having the same spatial resolution and level of detail.
In the literature, even one or several of these constraints have been shown problematic for carrying out a registration method (see Section \ref{sec:related-work}).
To the best of our knowledge, there is currently no solution able to achieve the registration between airborne LiDAR and optical imagery under such constraints altogether.
As a matter of fact, the proposed registration method has been evaluated according to its own quality, before and after the registration. 
Indeed, it is not compared with existing methods because they were not designed to address the considered context.
Nevertheless, we reconsidered the subjective accuracy suggestion related to a sub-pixel level of accuracy for a registration (see \ref{sssec:accuracy}).
Instead, we rely on an objectively quantitative accuracy which is that, if the resulting spatial discrepancy less than 50 cm, then the registration will be considered accurate.
In this regard, the proposed registration method has achieved a highly desirable accuracy.

The proposed method can be summarized as follows.
First, a coarse feature-based registration is carried out based on the extraction and matching of building candidates on the two datasets, reducing significantly the spatial shift between them. 
Here, it should also be noted that this building-based  approach certainly does not limit the usability and versatility of our method, since urban scenes with buildings (even very sparse) are available most of the time \cite{Vlahov2002}.
Then, a fine registration based on the maximization of MI or NCMI (both measures have been performed separately) is carried out to determine the optimal camera pose, granting the datasets to be precisely aligned. 
It involves a process of super-resolution of LiDAR data to generate high-resolution images of altitude and intensity values. 
This approach neutralizes the difference of spatial resolution and level of detail between datasets, enabling the MI-based and NCMI-based fine registration.
The fine registration also involves in dividing the considered area into many equal patches. 
For each patch a local transformation model is estimated. 
This approach allows reducing significantly the computational cost of the fine registration. 
Lastly, a smoothing of the patch-based transformation models is carried out to resolve the conflicts and discontinuities between them. It involves an IDW average of camera pose parameters from neighboring patches.

As one can realize, many elements of the proposed method are intended as the solution to the challenges associated with the considered context. 
First, in order to address the spatial {shift} between datasets caused by the differences of points of view and fields of view, a coarse registration is carried out. 
It relies on using buildings as primitives, which is a relevant choice of primitive considering the low density of airborne LiDAR point cloud around vertical surfaces. 
Then, the differences of spatial resolution and level of detail  between datasets have been dealt with by the SR approach. 
An area-based fine registration using MI or NCMI measurement is carried out to finely tune the optimal local transformation model. 
Overall, as highlighted by the comprehensive spatial discrepancy assessments, the proposed method has achieved a very high registration accuracy. 
It is especially desirable when taking into account the difficulties of the considered context, and the horizontal accuracy of the datasets.

It is suggested that only one registration approach is not sufficient to register the data accurately from heterogeneous sensors, even when they are rigidly fixed to the same platform \cite{brell2016improving}. 
In this paper, we proposed a coarse-to-fine registration method consisting of two steps of registration.
It reinforces the relevance of a coarse-to-fine approach for registering an optical aerial/satellite imagery with an airborne LiDAR dataset.
Nevertheless, it can be anticipated that the proposed approach could have limitations to operate in an environment lacking of man-made objects providing reliable primitives, such as forest and desert areas. 
Thus, a study on the relevance and reliability of primitives found on these environments is necessary for an effective solution therein.
However, if we could carry out the coarse registration with manual control points, the proposed subsequent fine registration would not be limited and can be well carried out on these scenes.
Finally, with these promising results, the reported research has established a basis for a comprehensive fusion of aerial/satellite optical imagery and airborne LiDAR data in future researches.


%

\section*{Acknowledgment}
The authors would like to thank Centre G\'{e}oStat, Universit\'{e} Laval, as well as Qu\'{e}bec City, Communaut\'{e} M\'{e}tropolitaine de Qu\'{e}bec (QC, Canada), and Centre National d'Etudes Spatiales (France) for providing the datasets used in this work.
We also would like to thank Dr. Hongchao Ma, School of Remote Sensing and Information Engineering, Wuhan University, China for his/her help on the usage of the check-pair line evaluation.

\ifCLASSOPTIONcaptionsoff
  \newpage
\fi



\bibliographystyle{IEEEtran}
\bibliography{IEEEabrv,reference_abbr.bib}
\end{document}